\documentclass[12pt, draftclsnofoot, onecolumn]{IEEEtran}
\usepackage{amssymb}
\usepackage[cmex10]{amsmath}
\usepackage{stfloats}
\usepackage{graphicx}
\usepackage{subfigure}
\usepackage{tabularx}
\usepackage{epsfig,epsf,color,balance,cite}
\usepackage{verbatim}
\usepackage{url}
\usepackage{bm}
\usepackage{mathrsfs}
\usepackage{amsthm}
\usepackage{mathtools}

\newtheorem{theorem}{Theorem}

\usepackage{algorithm}
\usepackage{algpseudocode}
\usepackage{graphicx}
\usepackage{epstopdf}
\begin{document}

\title{A Note on Fourier-Motzkin Elimination with Three Eliminating Variables}

\author{
	\IEEEauthorblockN{Hao Xu, \emph{Member, IEEE}\IEEEauthorrefmark{0},    
		Kai-Kit Wong, \emph{Fellow, IEEE}\IEEEauthorrefmark{0},
		and
		Giuseppe Caire, \emph{Fellow, IEEE}\IEEEauthorrefmark{0}
	}
	\thanks{
		This work was supported by the European Union's Horizon 2020 Research and Innovation Programme under Marie Skłodowska-Curie Grant No. 101024636 and the Alexander von Humboldt Foundation.
		
		H. Xu and K.-K. Wong are with the Department of Electronic and Electrical Engineering, University College London, London WC1E 7JE, UK (e-mail: hao.xu@ucl.ac.uk; kai-kit.wong@ucl.ac.uk).
		
		G. Caire is with the Faculty of Electrical Engineering and Computer Science at the Technical University of Berlin, 10587 Berlin, Germany (e-mail: caire@tu-berlin.de).
	}
}

\maketitle

\IEEEpeerreviewmaketitle

\vspace{-5em}
\section{Introduction}
\label{introduction}

In this note, we give Theorem~\ref{theorem_FM}, which arises from the capacity analysis for a multiple access wiretap (MAC-WT) channel with $K$ users, and prove it for the $K=3$ case \cite{9174164, xu2022achievable, xu2022achievable2}.
This theorem is given first in our work \cite[Lemma~$7$]{xu2022achievable} (special case with $K=2$) and then \cite[Lemma~$1$]{xu2022achievable2} (with general $K$).
For the simple system with a small $K$, e.g., $K = 1$ or $K = 2$, by eliminating $R_k^{\text g}$ in (\ref{region_FM1}) using the Fourier-Motzkin procedure \cite[Appendix D]{el2011network}, it can be easily proven that (\ref{rate_region0}) is the projection of (\ref{region_FM1}) onto the hyperplane $\{ R_k^{\text g} = 0, \forall k \in {\cal K}\}$.
Theorem~\ref{theorem_FM} can thus be proven for these simple cases.
However, when $K$ increases, the number of inequalities generated in the elimination procedure grows very quickly (doubly exponentially).
It thus becomes unmanageable to prove Theorem~\ref{theorem_FM} by directly using the Fourier-Motzkin procedure.
Besides the great complexity, another disadvantage of this strategy is that it works only if $K$ is given.
Obviously, this makes the strategy inappropriate for the proof of Theorem~\ref{theorem_FM} since it is a general result for any $K$.

In this note, we prove Theorem~\ref{theorem_FM} for the $K = 3$ case via eliminating $R_1^{\text g}$, $R_2^{\text g}$, and $R_3^{\text g}$ one by one.
From (\ref{region_FM1}) we first get $4$ upper bounds and $5$ lower bounds on $R_1^{\text g}$.
By pairing up these lower and upper bounds, we get $20$ inequalities, based on which $8$ upper bounds and $7$ lower bounds on $R_2^{\text g}$ are obtained.
We then eliminate $R_2^{\text g}$ and get $56$ inequalities, in which most of them are redundant.
Neglecting the redundant inequalities, we further get $9$ upper bounds and $7$ lower bounds on $R_3^{\text g}$, and $63$ inequalities by pairing them up.
Neglecting the redundant terms, we show that Theorem~\ref{theorem_FM} in this $K = 3$ case can be directly proven by Fourier-Motzkin elimination.

We can see from this note that even when $K = 3$, it is very difficult to project (\ref{region_FM1}) onto hyperplane $\{ R_1^{\text g} = 0, R_2^{\text g} = 0, R_3^{\text g} = 0\}$ and get (\ref{rate_region0}) since we have to consider so many inequalities and prove that most of them are redundant.
Obviously, when $K$ is large, it is impossible to prove Theorem~\ref{theorem_FM} by directly using Fourier-Motzkin elimination.
Therefore, we hope to circumvent this brute-force way and strictly prove Theorem~\ref{theorem_FM} for any $K$.
We provide the general proof of this theorem in \cite[Appendix A]{xu2022achievable2}.
Note that the redundancy of an inequality can be proven by showing that: 1) there exists at least one non-redundant inequality tighter than this one; 2) this inequality can be obtained by combining different inequalities; 3) a bound tighter than this inequality can be obtained by combining different inequalities, etc.
As for a specific inequality, its redundancy proof may not be unique.
For example, we may obtain this inequality or a tighter bound by different combinations.
Actually, most of the redundant inequalities admit different redundancy proofs.
We show this in the elimination procedure in this note.
But due to space limitation and the massive work, we only give some examples in e12, g12, m19, n15, q19, s19, and t19.
To generally prove Theorem~\ref{theorem_FM}, we have to draw one rule with which the redundancy of all redundant inequalities can be proven.
This makes the general proof quite challenging.

In this note, we use calligraphic capital letters to denote sets, e.g., ${\cal K} = \{1, \cdots, K \}$, calligraphic subscript to denote the set of elements whose indexes take values from the subscript set, e.g., $X_{\cal K} = \{X_k, \forall k \in {\cal K}\}$, and ${\bar {\cal S}}$ to denote the complementary set of ${\cal S}$.
For brevity, we use $\rightarrow$ or $\leftarrow$ to indicate `causes' or `results in'.
Please refer to references \cite{9174164}, \cite{xu2022achievable}, or \cite{xu2022achievable2} for the physical meanings of all the variables.

\section{Problem Formulation}
\label{section1}

\begin{theorem}\label{theorem_FM}
	\cite[Lemma~$1$]{xu2022achievable2}
	Let $(X_{\cal K}, Y, Z) \sim \prod_{k=1}^K p(x_k) p(y,z| x_{\cal K})$.
	If $I(X_{\cal S}; Y| X_{\bar {\cal S}}) \geq I(X_{\cal S}; Z),$ $\forall {\cal S} \subseteq {\cal K}$, then, for any rate tuple $(R_1^{\text s}, R_1^{\text o},$ $\cdots, R_K^{\text s}, R_K^{\text o})$ satisfying  
	\begin{align}\label{rate_region0}
	\sum_{k \in \cal S} R_k^{\text s} + \sum_{k \in {\cal S}\setminus {\cal S}_1} R_k^{\text o} \leq & I(X_{\cal S}; Y| X_{\bar {\cal S}}) - I(X_{{\cal S}_1}; Z), \forall {\cal S} \subseteq {\cal K},~ {{\cal S}_1} \subseteq \cal S,
	\end{align}
	there exists $(R_1^{\text g}, \cdots, R_K^{\text g})$ such that
	\begin{equation}\label{region_FM1}
	\left\{\!\!\!
	\begin{array}{ll}
	R_k^{\text g} \geq 0, ~\forall~ k \in {\cal K}, \\
	\sum\limits_{k \in {\cal S}} (R_k^{\text s} + R_k^{\text o} + R_k^{\text g}) \leq I(X_{\cal S}; Y| X_{\bar {\cal S}}), \forall {\cal S} \subseteq {\cal K}, \\
	\sum\limits_{k \in {\cal S}} (R_k^{\text o} + R_k^{\text g}) \geq I(X_{\cal S}; Z), \forall {\cal S} \subseteq {\cal K}.
	\end{array} \right.
	\end{equation}
\end{theorem}

As explained in the introduction, if $K$ is small, Theorem~\ref{theorem_FM} can be easily proven by directly using the Fourier-Motzkin procedure \cite[Appendix D]{el2011network} to eliminate $R_k^{\text g}$ in (\ref{region_FM1}) and get (\ref{rate_region0}).
If $K = 3$, (\ref{region_FM1}) can be written in detail as follows
\begin{align}
& R_1^{\text s} + R_1^{\text o} + R_1^{\text g} \leq I(X_1; Y| X_2, X_3), \label{region_before_elimi_1}\\
& R_2^{\text s} + R_2^{\text o} + R_2^{\text g} \leq I(X_2; Y| X_1, X_3),\label{region_before_elimi_2}\\
& R_3^{\text s} + R_3^{\text o} + R_3^{\text g} \leq I(X_3; Y| X_1, X_2),\label{region_before_elimi_3}\\
& R_1^{\text s} + R_1^{\text o} + R_1^{\text g} + R_2^{\text s} + R_2^{\text o} + R_2^{\text g} \leq I(X_1, X_2; Y| X_3),\label{region_before_elimi_4}\\
& R_1^{\text s} + R_1^{\text o} + R_1^{\text g} + R_3^{\text s} + R_3^{\text o} + R_3^{\text g} \leq I(X_1, X_3; Y| X_2),\label{region_before_elimi_5}\\
& R_2^{\text s} + R_2^{\text o} + R_2^{\text g} + R_3^{\text s} + R_3^{\text o} + R_3^{\text g} \leq I(X_2, X_3; Y| X_1),\label{region_before_elimi_6}\\
& R_1^{\text s} + R_1^{\text o} + R_1^{\text g} + R_2^{\text s} + R_2^{\text o} + R_2^{\text g} + R_3^{\text s} + R_3^{\text o} + R_3^{\text g} \leq I(X_1, X_2, X_3; Y),\label{region_before_elimi_7}\\
& R_1^{\text g} \geq 0,\label{region_before_elimi_8}\\
& R_2^{\text g} \geq 0,\label{region_before_elimi_9}\\
& R_3^{\text g} \geq 0,\label{region_before_elimi_10}\\
& R_1^{\text o} + R_1^{\text g} \geq I(X_1; Z),\label{region_before_elimi_11}\\
& R_2^{\text o} + R_2^{\text g} \geq I(X_2; Z),\label{region_before_elimi_12}\\
& R_3^{\text o} + R_3^{\text g} \geq I(X_3; Z),\label{region_before_elimi_13}\\
& R_1^{\text o} + R_1^{\text g} + R_2^{\text o} + R_2^{\text g} \geq I(X_1, X_2; Z),\label{region_before_elimi_14}\\
& R_1^{\text o} + R_1^{\text g} + R_3^{\text o} + R_3^{\text g} \geq I(X_1, X_3; Z),\label{region_before_elimi_15}\\
& R_2^{\text o} + R_2^{\text g} + R_3^{\text o} + R_3^{\text g} \geq I(X_2, X_3; Z),\label{region_before_elimi_16}\\
& R_1^{\text o} + R_1^{\text g} + R_2^{\text o} + R_2^{\text g} + R_3^{\text o} + R_3^{\text g} \geq I(X_1, X_2, X_3; Z).\label{region_before_elimi_17}
\end{align}
In the next section, we prove Theorem~\ref{theorem_FM} for the $K=3$ case and show the great complexity.
Note that the general poof of Theorem~\ref{theorem_FM} for any $K$ has been provided in our work \cite{xu2022achievable2}.

\section{Proof of Theorem~\ref{theorem_FM} Using Fourier-Motzkin Elimination}

In this section, we prove Theorem~\ref{theorem_FM} via eliminating $R_1^{\text g}$, $R_2^{\text g}$, and $R_3^{\text g}$ one by one, and show that even when $K = 3$, it is very difficult to project (\ref{region_FM1}) onto hyperplane $\{ R_1^{\text g} = 0, R_2^{\text g} = 0, R_3^{\text g} = 0\}$ since we have to consider so many inequalities and prove that most of them are redundant.
For easy reading, the inequalities necessary for the next step elimination are in black color, the efficient projected inequalities are in \textcolor[rgb]{0.00,0.59,0.00}{green color}, and all redundant inequalities are in \textcolor[rgb]{1.00,0.00,0.00}{red color}.
Besides, we explain why it is redundant after each redundant inequality in \textcolor[rgb]{0.00,0.07,1.00}{blue fonts}.

\subsection{Elimination of $R_1^{g}$}

We first eliminate $R_1^{g}$.
From (\ref{region_before_elimi_1}) $\sim$ (\ref{region_before_elimi_17}), we get $4$ upper bounds on $R_1^{\text g}$
\begin{align}
(\ref{region_before_elimi_1}) \rightarrow & {\text a}. \quad R_1^{\text g} \leq I(X_1; Y| X_2, X_3) - (R_1^{\text s} + R_1^{\text o}), \nonumber\\
(\ref{region_before_elimi_4}) \rightarrow & {\text b}. \quad R_1^{\text g} \leq I(X_1, X_2; Y| X_3) - (R_1^{\text s} + R_1^{\text o} + R_2^{\text s} + R_2^{\text o} + R_2^{\text g}), \nonumber\\
(\ref{region_before_elimi_5}) \rightarrow & {\text c}. \quad R_1^{\text g} \leq I(X_1, X_3; Y| X_2) - (R_1^{\text s} + R_1^{\text o} + R_3^{\text s} + R_3^{\text o} + R_3^{\text g}), \nonumber\\
(\ref{region_before_elimi_7}) \rightarrow & {\text d}. \quad R_1^{\text g} \leq I(X_1, X_2, X_3; Y) - (R_1^{\text s} + R_1^{\text o} + R_2^{\text s} + R_2^{\text o} + R_2^{\text g} + R_3^{\text s} + R_3^{\text o} + R_3^{\text g}),\nonumber
\end{align}
and $5$ lower bounds on $R_1^{\text g}$
\begin{align}
(\ref{region_before_elimi_8}) \rightarrow & {\text 1}. \quad R_1^{\text g} \geq 0, \nonumber\\
(\ref{region_before_elimi_11}) \rightarrow & {\text 2}. \quad R_1^{\text g} \geq I(X_1; Z) - R_1^{\text o}, \nonumber\\
(\ref{region_before_elimi_14}) \rightarrow & {\text 3}. \quad R_1^{\text g} \geq I(X_1, X_2; Z) - ( R_1^{\text o} + R_2^{\text o} + R_2^{\text g}), \nonumber\\
(\ref{region_before_elimi_15}) \rightarrow & {\text 4}. \quad R_1^{\text g} \geq I(X_1, X_3; Z) - ( R_1^{\text o} + R_3^{\text o} + R_3^{\text g}), \nonumber\\
(\ref{region_before_elimi_17}) \rightarrow & {\text 5}. \quad R_1^{\text g} \geq I(X_1, X_2, X_3; Z) - ( R_1^{\text o} + R_2^{\text o} + R_2^{\text g} + R_3^{\text o} + R_3^{\text g}).\nonumber
\end{align}
For convenience, letters and numbers are respectively used to index the upper and lower bounds.
The combination of a letter and a number, e.g., `a1', indexes the inequality resulted from comparing upper bound `a' with lower bound `1'.

Comparing the upper bounds ${\text a} \sim {\text d}$ and lower bounds $1 \sim 5$ on $R_1^{\text g}$ given above, we get
\begin{align}
& \textcolor[rgb]{0.00,0.59,0.00}{{\text {a1}}. \quad R_1^{\text s} + R_1^{\text o} \leq I(X_1; Y| X_2, X_3),} \nonumber\\
& \textcolor[rgb]{0.00,0.59,0.00}{{\text {a2}}. \quad R_1^{\text s} \leq I(X_1; Y| X_2, X_3) - I(X_1; Z),} \nonumber\\
& {\text {a3}}. \quad R_1^{\text s} - ( R_2^{\text o} + R_2^{\text g})  \leq I(X_1; Y| X_2, X_3) - I(X_1, X_2; Z), \nonumber\\
& {\text {a4}}. \quad R_1^{\text s} - ( R_3^{\text o} + R_3^{\text g}) \leq I(X_1; Y| X_2, X_3) - I(X_1, X_3; Z), \nonumber\\
& {\text {a5}}. \quad R_1^{\text s} - (R_2^{\text o} + R_2^{\text g} + R_3^{\text o} + R_3^{\text g}) \leq I(X_1; Y| X_2, X_3) - I(X_1, X_2, X_3; Z), \nonumber\\
& {\text {b1}}. \quad R_1^{\text s} + R_1^{\text o} + R_2^{\text s} + R_2^{\text o} + R_2^{\text g} \leq I(X_1, X_2; Y| X_3), \nonumber\\
& {\text {b2}}. \quad R_1^{\text s} + R_2^{\text s} + R_2^{\text o} + R_2^{\text g} \leq I(X_1, X_2; Y| X_3) - I(X_1; Z), \nonumber\\
& \textcolor[rgb]{0.00,0.59,0.00}{{\text {b3}}. \quad R_1^{\text s} + R_2^{\text s} \leq I(X_1, X_2; Y| X_3) - I(X_1, X_2; Z),} \nonumber\\
& {\text {b4}}. \quad R_1^{\text s} + R_2^{\text s} + R_2^{\text o} + R_2^{\text g} - ( R_3^{\text o} + R_3^{\text g}) \leq I(X_1, X_2; Y| X_3) - I(X_1, X_3; Z), \nonumber\\
& {\text {b5}}. \quad R_1^{\text s} + R_2^{\text s} - (R_3^{\text o} + R_3^{\text g}) \leq I(X_1, X_2; Y| X_3) - I(X_1, X_2, X_3; Z), \nonumber\\
& {\text {c1}}. \quad R_1^{\text s} + R_1^{\text o} + R_3^{\text s} + R_3^{\text o} + R_3^{\text g} \leq I(X_1, X_3; Y| X_2), \nonumber\\
& {\text {c2}}. \quad R_1^{\text s} + R_3^{\text s} + R_3^{\text o} + R_3^{\text g} \leq I(X_1, X_3; Y| X_2) - I(X_1; Z), \nonumber\\
& {\text {c3}}. \quad R_1^{\text s} + R_3^{\text s} + R_3^{\text o} + R_3^{\text g} - ( R_2^{\text o} + R_2^{\text g}) \leq I(X_1, X_3; Y| X_2) - I(X_1, X_2; Z), \nonumber\\
& \textcolor[rgb]{0.00,0.59,0.00}{{\text {c4}}. \quad R_1^{\text s} + R_3^{\text s} \leq I(X_1, X_3; Y| X_2) - I(X_1, X_3; Z),} \nonumber\\
& {\text {c5}}. \quad R_1^{\text s} + R_3^{\text s} - (R_2^{\text o} + R_2^{\text g}) \leq I(X_1, X_3; Y| X_2) - I(X_1, X_2, X_3; Z), \nonumber\\
& {\text {d1}}. \quad R_1^{\text s} + R_1^{\text o} + R_2^{\text s} + R_2^{\text o} + R_2^{\text g} + R_3^{\text s} + R_3^{\text o} + R_3^{\text g} \leq I(X_1, X_2, X_3; Y), \nonumber\\
& {\text {d2}}. \quad R_1^{\text s} + R_2^{\text s} + R_2^{\text o} + R_2^{\text g} + R_3^{\text s} + R_3^{\text o} + R_3^{\text g} \leq I(X_1, X_2, X_3; Y) - I(X_1; Z), \nonumber\\
& {\text {d3}}. \quad R_1^{\text s} + R_2^{\text s} + R_3^{\text s} + R_3^{\text o} + R_3^{\text g} \leq I(X_1, X_2, X_3; Y) - I(X_1, X_2; Z), \nonumber\\
& {\text {d4}}. \quad R_1^{\text s} + R_2^{\text s} + R_2^{\text o} + R_2^{\text g} + R_3^{\text s} \leq I(X_1, X_2, X_3; Y) - I(X_1, X_3; Z), \nonumber\\
& \textcolor[rgb]{0.00,0.59,0.00}{{\text {d5}}. \quad R_1^{\text s} + R_2^{\text s} + R_3^{\text s} \leq I(X_1, X_2, X_3; Y) - I(X_1, X_2, X_3; Z),} \nonumber
\end{align}
The inequalities in green color, e.g., `a1', `a2', `b3', `c4', and `d5', are efficient, i.e., they will appear in the final projected inequality system (\ref{rate_region0}) (see (\ref{projected_system}) at the end of this note for detail).
The other inequalities in black color contain $R_2^{\text g}$ or $R_3^{\text g}$.
Hence, they are still useful for the following elimination procedure.

\subsection{Elimination of $R_2^{g}$}

From (\ref{region_before_elimi_1}) $\sim$ (\ref{region_before_elimi_17}) and the inequalities resulted from eliminating $R_1^{\text g}$ in the previous subsection, we get $8$ upper bounds on $R_2^{\text g}$
\begin{align}
(\ref{region_before_elimi_2}) \rightarrow & {\text e}. \quad R_2^{\text g} \leq I(X_2; Y| X_1, X_3) - (R_2^{\text s} + R_2^{\text o}), \nonumber\\
(\ref{region_before_elimi_6}) \rightarrow & {\text f}. \quad R_2^{\text g} \leq I(X_2, X_3; Y| X_1) - (R_2^{\text s} + R_2^{\text o} + R_3^{\text s} + R_3^{\text o} + R_3^{\text g}), \nonumber\\
{\text {b1}} \rightarrow & {\text g}. \quad R_2^{\text g} \leq I(X_1, X_2; Y| X_3) - (R_1^{\text s} + R_1^{\text o} + R_2^{\text s} + R_2^{\text o}), \nonumber\\
{\text {b2}} \rightarrow & {\text h}. \quad R_2^{\text g} \leq I(X_1, X_2; Y| X_3) - I(X_1; Z) - (R_1^{\text s} + R_2^{\text s} + R_2^{\text o}),\nonumber\\
{\text {b4}} \rightarrow & {\text i}. \quad R_2^{\text g} \leq I(X_1, X_2; Y| X_3) - I(X_1, X_3; Z) - (R_1^{\text s} + R_2^{\text s} + R_2^{\text o}) + ( R_3^{\text o} + R_3^{\text g}), \nonumber\\
{\text {d1}} \rightarrow & {\text j}. \quad R_2^{\text g} \leq I(X_1, X_2, X_3; Y) - (R_1^{\text s} + R_1^{\text o} + R_2^{\text s} + R_2^{\text o} + R_3^{\text s} + R_3^{\text o} + R_3^{\text g}),\nonumber\\
{\text {d2}} \rightarrow & {\text k}. \quad R_2^{\text g} \leq I(X_1, X_2, X_3; Y) - I(X_1; Z) - (R_1^{\text s} + R_2^{\text s} + R_2^{\text o} + R_3^{\text s} + R_3^{\text o} + R_3^{\text g}), \nonumber\\
{\text {d4}} \rightarrow & {\text L}. \quad R_2^{\text g} \leq I(X_1, X_2, X_3; Y) - I(X_1, X_3; Z) - (R_1^{\text s} + R_2^{\text s} + R_2^{\text o} + R_3^{\text s}),\nonumber
\end{align}
and $7$ lower bounds on $R_2^{\text g}$
\begin{align}
(\ref{region_before_elimi_9}) \rightarrow & {\text 6}. \quad R_2^{\text g} \geq 0, \nonumber\\
(\ref{region_before_elimi_12}) \rightarrow & {\text 7}. \quad R_2^{\text g} \geq I(X_2; Z) - R_2^{\text o}, \nonumber\\
(\ref{region_before_elimi_16}) \rightarrow & {\text 8}. \quad R_2^{\text g} \geq I(X_2, X_3; Z) - (R_2^{\text o} + R_3^{\text o} + R_3^{\text g}), \nonumber\\
{\text {a3}} \rightarrow & {\text 9}. \quad R_2^{\text g} \geq - I(X_1; Y| X_2, X_3) + I(X_1, X_2; Z) + R_1^{\text s} - R_2^{\text o},\nonumber\\
{\text {a5}} \rightarrow & {\text 10}. \quad R_2^{\text g} \geq - I(X_1; Y| X_2, X_3) + I(X_1, X_2, X_3; Z) + R_1^{\text s} - (R_2^{\text o} + R_3^{\text o} + R_3^{\text g}), \nonumber\\
{\text {c3}} \rightarrow & {\text 11}. \quad R_2^{\text g} \geq - I(X_1, X_3; Y| X_2) + I(X_1, X_2; Z) + R_1^{\text s} + R_3^{\text s} + R_3^{\text o} + R_3^{\text g} - R_2^{\text o},\nonumber\\
{\text {c5}} \rightarrow & {\text 12}. \quad R_2^{\text g} \geq - I(X_1, X_3; Y| X_2) + I(X_1, X_2, X_3; Z) + R_1^{\text s} + R_3^{\text s} - R_2^{\text o}, \nonumber
\end{align}

Comparing the upper bounds ${\text e} \sim {\text L}$ and lower bounds $6 \sim 12$ on $R_2^{\text g}$ given above, we get
\begin{align}
& \textcolor[rgb]{0.00,0.59,0.00}{{\text {e6}}. \quad R_2^{\text s} + R_2^{\text o} \leq I(X_2; Y| X_1, X_3),} \nonumber\\
& \textcolor[rgb]{0.00,0.59,0.00}{{\text {e7}}. \quad R_2^{\text s} \leq I(X_2; Y| X_1, X_3) - I(X_2; Z),} \nonumber\\
& {\text {e8}}. \quad R_2^{\text s} - (R_3^{\text o} + R_3^{\text g}) \leq I(X_2; Y| X_1, X_3) - I(X_2, X_3; Z), \nonumber\\
& \textcolor[rgb]{1.00,0.00,0.00}{{\text {e9}}. \quad R_1^{\text s} + R_2^{\text s} \leq I(X_2; Y| X_1, X_3) + I(X_1; Y| X_2, X_3) - I(X_1, X_2; Z),} \nonumber\\
& \textcolor[rgb]{0.00,0.07,1.00}{\quad\quad R_1^{\text s} + R_2^{\text s} \leq I(X_1, X_2; Y| X_3) - I(X_1, X_2; Z)  \leftarrow {\text {b3}}} \nonumber\\
& \textcolor[rgb]{0.00,0.07,1.00}{\quad\quad\quad\quad\quad\quad\! \leq I(X_2; Y| X_1, X_3) + I(X_1; Y| X_2, X_3) - I(X_1, X_2; Z)} \nonumber\\
& \textcolor[rgb]{1.00,0.00,0.00}{{\text {e10}}. \quad R_1^{\text s} + R_2^{\text s} - ( R_3^{\text o} + R_3^{\text g}) \leq I(X_2; Y| X_1, X_3) + I(X_1; Y| X_2, X_3) - I(X_1, X_2, X_3; Z),} \nonumber\\
& \textcolor[rgb]{0.00,0.07,1.00}{\quad\quad\quad R_1^{\text s} + R_2^{\text s} - (R_3^{\text o} + R_3^{\text g}) \leq I(X_1, X_2; Y| X_3) - I(X_1, X_2, X_3; Z) \leftarrow {\text {b5}}}\nonumber\\
& \textcolor[rgb]{0.00,0.07,1.00}{\quad\quad\quad\quad\quad\quad\quad\quad\quad\quad\quad\quad\; \leq I(X_2; Y| X_1, X_3) + I(X_1; Y| X_2, X_3) - I(X_1, X_2, X_3; Z)} \nonumber\\
& \textcolor[rgb]{1.00,0.00,0.00}{{\text {e11}}. \quad R_1^{\text s} + R_2^{\text s} + R_3^{\text s} + R_3^{\text o} + R_3^{\text g} \leq I(X_2; Y| X_1, X_3) + I(X_1, X_3; Y| X_2) - I(X_1, X_2; Z),} \nonumber\\
& \textcolor[rgb]{0.00,0.07,1.00}{\quad\quad\quad R_1^{\text s} + R_2^{\text s} + R_3^{\text s} + R_3^{\text o} + R_3^{\text g} \leq I(X_1, X_2, X_3; Y) - I(X_1, X_2; Z) \leftarrow {\text {d3}}}\nonumber\\
& \textcolor[rgb]{0.00,0.07,1.00}{\quad\quad\quad\quad\quad\quad\quad\quad\quad\quad\quad\quad\quad\quad\! \leq I(X_2; Y| X_1, X_3) + I(X_1, X_3; Y| X_2) - I(X_1, X_2; Z)} \nonumber\\
& \textcolor[rgb]{1.00,0.00,0.00}{{\text {e12}}. \quad R_1^{\text s} + R_2^{\text s} + R_3^{\text s} \leq I(X_2; Y| X_1, X_3) + I(X_1, X_3; Y| X_2) - I(X_1, X_2, X_3; Z),} \nonumber\\
& \textcolor[rgb]{0.00,0.07,1.00}{\quad\quad\quad R_1^{\text s} + R_2^{\text s} + R_3^{\text s} \leq I(X_1, X_2, X_3; Y) - I(X_1, X_2, X_3; Z) \leftarrow {\text {d5}}}\nonumber\\
& \textcolor[rgb]{0.00,0.07,1.00}{\quad\quad\quad\quad\quad\quad\quad\quad\quad\; \leq I(X_2; Y| X_1, X_3) + I(X_1, X_3; Y| X_2) - I(X_1, X_2, X_3; Z)} \nonumber\\
& \textcolor[rgb]{0.00,0.07,1.00}{\quad\quad\quad {\text{or}}} \nonumber\\
& \textcolor[rgb]{0.00,0.07,1.00}{\quad\quad\quad (R_2^{\text s}) + (R_1^{\text s} + R_3^{\text s}) }\nonumber\\
& \textcolor[rgb]{0.00,0.07,1.00}{\quad\quad \leq I(X_2; Y| X_1, X_3) - I(X_2; Z) + I(X_1, X_3; Y| X_2) - I(X_1, X_3; Z) \leftarrow {\text {e7 + c4}}}\nonumber\\
& \textcolor[rgb]{0.00,0.07,1.00}{\quad\quad \leq I(X_2; Y| X_1, X_3) + I(X_1, X_3; Y| X_2) - I(X_1, X_2, X_3; Z) }\nonumber
\end{align}
\begin{align}
& {\text {f6}}. \quad R_2^{\text s} + R_2^{\text o} + R_3^{\text s} + R_3^{\text o} + R_3^{\text g} \leq I(X_2, X_3; Y| X_1), \nonumber\\
& {\text {f7}}. \quad R_2^{\text s} + R_3^{\text s} + R_3^{\text o} + R_3^{\text g} \leq I(X_2, X_3; Y| X_1) - I(X_2; Z), \nonumber\\
& \textcolor[rgb]{0.00,0.59,0.00}{{\text {f8}}. \quad R_2^{\text s} + R_3^{\text s} \leq I(X_2, X_3; Y| X_1) - I(X_2, X_3; Z),} \nonumber\\
& \textcolor[rgb]{1.00,0.00,0.00}{{\text {f9}}. \quad R_1^{\text s} + R_2^{\text s} + R_3^{\text s} + R_3^{\text o} + R_3^{\text g} \leq I(X_2, X_3; Y| X_1) + I(X_1; Y| X_2, X_3) - I(X_1, X_2; Z),} \nonumber\\
& \textcolor[rgb]{0.00,0.07,1.00}{\quad\quad R_1^{\text s} + R_2^{\text s} + R_3^{\text s} + R_3^{\text o} + R_3^{\text g} \leq I(X_1, X_2, X_3; Y) - I(X_1, X_2; Z)  \leftarrow {\text {d3}}}\nonumber\\
& \textcolor[rgb]{0.00,0.07,1.00}{\quad\quad\quad\quad\quad\quad\quad\quad\quad\quad\quad\quad\quad\! \leq I(X_2, X_3; Y| X_1) + I(X_1; Y| X_2, X_3) - I(X_1, X_2; Z)}\nonumber\\
& \textcolor[rgb]{1.00,0.00,0.00}{{\text {f10}}. \quad R_1^{\text s} + R_2^{\text s} + R_3^{\text s} \leq I(X_2, X_3; Y| X_1) + I(X_1; Y| X_2, X_3) - I(X_1, X_2, X_3; Z),} \nonumber\\
& \textcolor[rgb]{0.00,0.07,1.00}{\quad\quad\quad R_1^{\text s} + R_2^{\text s} + R_3^{\text s} \leq I(X_1, X_2, X_3; Y) - I(X_1, X_2, X_3; Z)  \leftarrow {\text {d5}}}\nonumber\\
& \textcolor[rgb]{0.00,0.07,1.00}{\quad\quad\quad\quad\quad\quad\quad\quad \leq I(X_2, X_3; Y| X_1) + I(X_1; Y| X_2, X_3) - I(X_1, X_2, X_3; Z)}\nonumber\\
& \textcolor[rgb]{1.00,0.00,0.00}{{\text {f11}}. \quad R_1^{\text s} + R_2^{\text s} + 2 R_3^{\text s} + 2 R_3^{\text o} + 2 R_3^{\text g} \leq I(X_2, X_3; Y| X_1) + I(X_1, X_3; Y| X_2) - I(X_1, X_2; Z),} \nonumber\\
& \textcolor[rgb]{0.00,0.07,1.00}{\quad\quad\quad (R_1^{\text s} + R_2^{\text s} + R_3^{\text s} + R_3^{\text o} + R_3^{\text g}) + (R_3^{\text s} + R_3^{\text o} + R_3^{\text g})}\nonumber\\
& \textcolor[rgb]{0.00,0.07,1.00}{\quad\quad \leq I(X_1, X_2, X_3; Y) - I(X_1, X_2; Z) + I(X_3; Y| X_1, X_2) \leftarrow {\text {d3 + (5)}}}\nonumber\\
& \textcolor[rgb]{0.00,0.07,1.00}{\quad\quad = I(X_2, X_3; Y| X_1) + I(X_1; Y) - I(X_1, X_2; Z) + I(X_3; Y| X_1, X_2)}\nonumber\\
& \textcolor[rgb]{0.00,0.07,1.00}{\quad\quad \leq I(X_2, X_3; Y| X_1) + I(X_1, X_3; Y| X_2) - I(X_1, X_2; Z)}\nonumber\\
& \textcolor[rgb]{1.00,0.00,0.00}{{\text {f12}}. \quad R_1^{\text s} + R_2^{\text s} + 2 R_3^{\text s} + R_3^{\text o} + R_3^{\text g} \leq I(X_2, X_3; Y| X_1) + I(X_1, X_3; Y| X_2) - I(X_1, X_2, X_3; Z),} \nonumber\\
& \textcolor[rgb]{0.00,0.07,1.00}{\quad\quad\quad (R_1^{\text s} + R_2^{\text s} + R_3^{\text s}) + (R_3^{\text s} + R_3^{\text o} + R_3^{\text g})}\nonumber\\
& \textcolor[rgb]{0.00,0.07,1.00}{\quad\quad \leq I(X_1, X_2, X_3; Y) - I(X_1, X_2, X_3; Z) + I(X_3; Y| X_1, X_2) \leftarrow {\text {d5 + (5)}}}\nonumber\\
& \textcolor[rgb]{0.00,0.07,1.00}{\quad\quad = I(X_1, X_3; Y| X_2) - I(X_1, X_2, X_3; Z) + I(X_2; Y) + I(X_3; Y| X_1, X_2)}\nonumber\\
& \textcolor[rgb]{0.00,0.07,1.00}{\quad\quad \leq I(X_1, X_3; Y| X_2) - I(X_1, X_2, X_3; Z) + I(X_2; Y| X_1) + I(X_3; Y| X_1, X_2)}\nonumber\\
& \textcolor[rgb]{0.00,0.07,1.00}{\quad\quad \leq I(X_1, X_3; Y| X_2) - I(X_1, X_2, X_3; Z) + I(X_2, X_3; Y| X_1)} \nonumber
\end{align}
\begin{align}
& \textcolor[rgb]{0.00,0.59,0.00}{{\text {g6}}. \quad R_1^{\text s} + R_1^{\text o} + R_2^{\text s} + R_2^{\text o} \leq I(X_1, X_2; Y| X_3),} \nonumber\\
& \textcolor[rgb]{0.00,0.59,0.00}{{\text {g7}}. \quad R_1^{\text s} + R_1^{\text o} + R_2^{\text s} \leq I(X_1, X_2; Y| X_3) - I(X_2; Z),} \nonumber\\
& {\text {g8}}. \quad R_1^{\text s} + R_1^{\text o} + R_2^{\text s} - (R_3^{\text o} + R_3^{\text g}) \leq I(X_1, X_2; Y| X_3) - I(X_2, X_3; Z), \nonumber\\
& \textcolor[rgb]{1.00,0.00,0.00}{{\text {g9}}. \quad 2 R_1^{\text s} + R_1^{\text o} + R_2^{\text s} \leq I(X_1, X_2; Y| X_3) + I(X_1; Y| X_2, X_3) - I(X_1, X_2; Z),} \nonumber\\
& \textcolor[rgb]{0.00,0.07,1.00}{\quad\quad\quad (R_1^{\text s} + R_2^{\text s}) + (R_1^{\text s} + R_1^{\text o}) }\nonumber\\
& \textcolor[rgb]{0.00,0.07,1.00}{\quad\quad \leq I(X_1, X_2; Y| X_3) - I(X_1, X_2; Z) + I(X_1; Y| X_2, X_3) \leftarrow {\text {b3 + a1}}}\nonumber\\
& \textcolor[rgb]{1.00,0.00,0.00}{{\text {g10}}. \quad 2 R_1^{\text s} + R_1^{\text o} + R_2^{\text s} - ( R_3^{\text o} + R_3^{\text g}) \leq I(X_1, X_2; Y| X_3) + I(X_1; Y| X_2, X_3) - I(X_1, X_2, X_3; Z),} \nonumber\\
& \textcolor[rgb]{0.00,0.07,1.00}{\quad\quad\quad (R_1^{\text s} + R_1^{\text o}) + [R_1^{\text s} + R_2^{\text s} - ( R_3^{\text o} + R_3^{\text g})]}\nonumber\\
& \textcolor[rgb]{0.00,0.07,1.00}{\quad\quad \leq I(X_1; Y| X_2, X_3) + I(X_1, X_2; Y| X_3) - I(X_1, X_2, X_3; Z) \leftarrow {\text {a1 + b5}}}\nonumber\\
& \textcolor[rgb]{1.00,0.00,0.00}{{\text {g11}}. \quad 2 R_1^{\text s} + R_1^{\text o} + R_2^{\text s} + R_3^{\text s} + R_3^{\text o} + R_3^{\text g} \leq I(X_1, X_2; Y| X_3) + I(X_1, X_3; Y| X_2) - I(X_1, X_2; Z),} \nonumber\\
& \textcolor[rgb]{0.00,0.07,1.00}{\quad\quad\quad (R_1^{\text s} + R_2^{\text s}) + ( R_1^{\text s} + R_1^{\text o} + R_3^{\text s} + R_3^{\text o} + R_3^{\text g})}\nonumber\\
& \textcolor[rgb]{0.00,0.07,1.00}{\quad\quad \leq I(X_1, X_2; Y| X_3) - I(X_1, X_2; Z) + I(X_1, X_3; Y| X_2) \leftarrow {\text {b3 + c1}}}\nonumber\\
& \textcolor[rgb]{1.00,0.00,0.00}{{\text {g12}}. \quad 2 R_1^{\text s} + R_1^{\text o} + R_2^{\text s} + R_3^{\text s} \leq I(X_1, X_2; Y| X_3) + I(X_1, X_3; Y| X_2) - I(X_1, X_2, X_3; Z),} \nonumber\\
& \textcolor[rgb]{0.00,0.07,1.00}{\quad\quad\quad (R_1^{\text s} + R_1^{\text o}) + ( R_1^{\text s} + R_2^{\text s} + R_3^{\text s})}\nonumber\\
& \textcolor[rgb]{0.00,0.07,1.00}{\quad\quad \leq I(X_1; Y| X_2, X_3) + I(X_1, X_2, X_3; Y) - I(X_1, X_2, X_3; Z) \leftarrow {\text {a1 + d5}}}\nonumber\\
& \textcolor[rgb]{0.00,0.07,1.00}{\quad\quad = I(X_1; Y| X_2, X_3) + I(X_2; Y) + I(X_1, X_3; Y| X_2) - I(X_1, X_2, X_3; Z)}\nonumber\\
& \textcolor[rgb]{0.00,0.07,1.00}{\quad\quad \leq I(X_1, X_2; Y| X_3) + I(X_1, X_3; Y| X_2) - I(X_1, X_2, X_3; Z)}\nonumber\\
& \textcolor[rgb]{0.00,0.07,1.00}{\quad\quad\quad {\text{or}} }\nonumber\\
& \textcolor[rgb]{0.00,0.07,1.00}{\quad\quad\quad (R_1^{\text s} + R_1^{\text o} + R_2^{\text s}) + ( R_1^{\text s} + R_3^{\text s})}\nonumber\\
& \textcolor[rgb]{0.00,0.07,1.00}{\quad\quad \leq I(X_1, X_2; Y| X_3) - I(X_2; Z) + I(X_1, X_3; Y| X_2) - I(X_1, X_3; Z) \leftarrow {\text {g7 + c4}}}\nonumber\\
& \textcolor[rgb]{0.00,0.07,1.00}{\quad\quad \leq I(X_1, X_2; Y| X_3) + I(X_1, X_3; Y| X_2) - I(X_1, X_2, X_3; Z)}\nonumber
\end{align}
\begin{align}
& \textcolor[rgb]{0.00,0.59,0.00}{{\text {h6}}. \quad R_1^{\text s} + R_2^{\text s} + R_2^{\text o} \leq I(X_1, X_2; Y| X_3) - I(X_1; Z),} \nonumber\\
& \textcolor[rgb]{1.00,0.00,0.00}{{\text {h7}}. \quad R_1^{\text s} + R_2^{\text s} \leq I(X_1, X_2; Y| X_3) - I(X_1; Z) - I(X_2; Z),} \nonumber\\
& \textcolor[rgb]{0.00,0.07,1.00}{\quad\quad R_1^{\text s} + R_2^{\text s} \leq I(X_1, X_2; Y| X_3) - I(X_1, X_2; Z) \leftarrow {\text {b3}}}\nonumber\\
& \textcolor[rgb]{0.00,0.07,1.00}{\quad\quad\quad\quad\quad\quad\! \leq I(X_1, X_2; Y| X_3) - I(X_1; Z) - I(X_2; Z)}\nonumber\\
& \textcolor[rgb]{1.00,0.00,0.00}{{\text {h8}}. \quad R_1^{\text s} + R_2^{\text s} - (R_3^{\text o} + R_3^{\text g}) \leq I(X_1, X_2; Y| X_3) - I(X_1; Z) - I(X_2, X_3; Z),} \nonumber\\
& \textcolor[rgb]{0.00,0.07,1.00}{\quad\quad R_1^{\text s} + R_2^{\text s} - (R_3^{\text o} + R_3^{\text g}) \leq I(X_1, X_2; Y| X_3) - I(X_1, X_2, X_3; Z) \leftarrow {\text {b5}}}\nonumber\\
& \textcolor[rgb]{0.00,0.07,1.00}{\quad\quad\quad\quad\quad\quad\quad\quad\quad\quad\quad\;\leq I(X_1, X_2; Y| X_3) - I(X_1; Z) - I(X_2, X_3; Z)}\nonumber\\
& \textcolor[rgb]{1.00,0.00,0.00}{{\text {h9}}. \quad 2 R_1^{\text s} + R_2^{\text s} \leq I(X_1, X_2; Y| X_3) - I(X_1; Z) + I(X_1; Y| X_2, X_3) - I(X_1, X_2; Z),} \nonumber\\
& \textcolor[rgb]{0.00,0.07,1.00}{\quad\quad R_1^{\text s} + \left( R_1^{\text s} + R_2^{\text s} \right) \leq I(X_1; Y| X_2, X_3) - I(X_1; Z) + I(X_1, X_2; Y| X_3) - I(X_1, X_2; Z) \leftarrow {\text {a2 + b3}}}\nonumber\\
& \textcolor[rgb]{1.00,0.00,0.00}{{\text {h10}}. \quad 2 R_1^{\text s} + R_2^{\text s} - ( R_3^{\text o} + R_3^{\text g}) \leq I(X_1, X_2; Y| X_3) - I(X_1; Z) + I(X_1; Y| X_2, X_3) - I(X_1, X_2, X_3; Z),} \nonumber\\
& \textcolor[rgb]{0.00,0.07,1.00}{\quad\quad\quad (R_1^{\text s}) + \left[ R_1^{\text s} + R_2^{\text s} - (R_3^{\text o} + R_3^{\text g}) \right]}\nonumber\\
& \textcolor[rgb]{0.00,0.07,1.00}{\quad\quad \leq I(X_1; Y| X_2, X_3) - I(X_1; Z) + I(X_1, X_2; Y| X_3) - I(X_1, X_2, X_3; Z) \leftarrow {\text {a2 + b5}}}\nonumber\\
& \textcolor[rgb]{1.00,0.00,0.00}{{\text {h11}}. \quad 2 R_1^{\text s} + R_2^{\text s} + R_3^{\text s} + R_3^{\text o} + R_3^{\text g} \leq I(X_1, X_2; Y| X_3) - I(X_1; Z) + I(X_1, X_3; Y| X_2) - I(X_1, X_2; Z),} \nonumber\\
& \textcolor[rgb]{0.00,0.07,1.00}{\quad\quad\quad (R_1^{\text s} + R_2^{\text s}) + ( R_1^{\text s} + R_3^{\text s} + R_3^{\text o} + R_3^{\text g})}\nonumber\\
& \textcolor[rgb]{0.00,0.07,1.00}{\quad\quad \leq I(X_1, X_2; Y| X_3) - I(X_1, X_2; Z) + I(X_1, X_3; Y| X_2) - I(X_1; Z) \leftarrow {\text {b3 + c2}}}\nonumber\\
& \textcolor[rgb]{1.00,0.00,0.00}{{\text {h12}}. \quad 2 R_1^{\text s} + R_2^{\text s} + R_3^{\text s} \leq I(X_1, X_2; Y| X_3) - I(X_1; Z) + I(X_1, X_3; Y| X_2) - I(X_1, X_2, X_3; Z),} \nonumber\\
& \textcolor[rgb]{0.00,0.07,1.00}{\quad\quad\quad (R_1^{\text s}) + ( R_1^{\text s} + R_2^{\text s} + R_3^{\text s})}\nonumber\\
& \textcolor[rgb]{0.00,0.07,1.00}{\quad\quad \leq I(X_1; Y| X_2, X_3) - I(X_1; Z) + I(X_1, X_2, X_3; Y) - I(X_1, X_2, X_3; Z) \leftarrow {\text {a2 + d5}}}\nonumber\\
& \textcolor[rgb]{0.00,0.07,1.00}{\quad\quad = I(X_1; Y| X_2, X_3) - I(X_1; Z) + I(X_2; Y) + I(X_1, X_3; Y| X_2) - I(X_1, X_2, X_3; Z)}\nonumber\\
& \textcolor[rgb]{0.00,0.07,1.00}{\quad\quad \leq I(X_1, X_2; Y| X_3) - I(X_1; Z) + I(X_1, X_3; Y| X_2) - I(X_1, X_2, X_3; Z)}\nonumber
\end{align}
\begin{align}
& {\text {i6}}. \quad R_1^{\text s} + R_2^{\text s} + R_2^{\text o} - (R_3^{\text o} + R_3^{\text g}) \leq I(X_1, X_2; Y| X_3) - I(X_1, X_3; Z), \nonumber\\
& \textcolor[rgb]{1.00,0.00,0.00}{{\text {i7}}. \quad R_1^{\text s} + R_2^{\text s} - (R_3^{\text o} + R_3^{\text g}) \leq I(X_1, X_2; Y| X_3) - I(X_1, X_3; Z) - I(X_2; Z),} \nonumber\\
& \textcolor[rgb]{0.00,0.07,1.00}{\quad\quad R_1^{\text s} + R_2^{\text s} - (R_3^{\text o} + R_3^{\text g}) \leq I(X_1, X_2; Y| X_3) - I(X_1, X_2, X_3; Z) \leftarrow {\text {b5}}}\nonumber\\
& \textcolor[rgb]{0.00,0.07,1.00}{\quad\quad\quad\quad\quad\quad\quad\quad\quad\quad\quad\; \leq I(X_1, X_2; Y| X_3) - I(X_1, X_3; Z) - I(X_2; Z)}\nonumber\\
& \textcolor[rgb]{1.00,0.00,0.00}{{\text {i8}}. \quad R_1^{\text s} + R_2^{\text s} - 2 (R_3^{\text o} + R_3^{\text g}) \leq I(X_1, X_2; Y| X_3) - I(X_1, X_3; Z) - I(X_2, X_3; Z),} \nonumber\\
& \textcolor[rgb]{0.00,0.07,1.00}{\quad\quad\quad [(R_1^{\text s} + R_2^{\text s}) - (R_3^{\text o} + R_3^{\text g})] - (R_3^{\text o} + R_3^{\text g}) }\nonumber\\
& \textcolor[rgb]{0.00,0.07,1.00}{\quad\quad \leq I(X_1, X_2; Y| X_3) - I(X_1, X_2, X_3; Z) - I(X_3; Z) \leftarrow {\text {b5 + (15)}}}\nonumber\\
& \textcolor[rgb]{0.00,0.07,1.00}{\quad\quad = I(X_1, X_2; Y| X_3) - I(X_1, X_3; Z) - I(X_2; Z| X_1, X_3) - I(X_3; Z)}\nonumber\\
& \textcolor[rgb]{0.00,0.07,1.00}{\quad\quad \leq I(X_1, X_2; Y| X_3) - I(X_1, X_3; Z) - I(X_2, X_3; Z)}\nonumber\\
& \textcolor[rgb]{1.00,0.00,0.00}{{\text {i9}}. \quad 2 R_1^{\text s} + R_2^{\text s} - (R_3^{\text o} + R_3^{\text g}) \leq I(X_1, X_2; Y| X_3) - I(X_1, X_3; Z) + I(X_1; Y| X_2, X_3) - I(X_1, X_2; Z),} \nonumber\\
& \textcolor[rgb]{0.00,0.07,1.00}{\quad\quad\quad ( R_1^{\text s} + R_2^{\text s}) + [R_1^{\text s} - (R_3^{\text o} + R_3^{\text g})] }\nonumber\\
& \textcolor[rgb]{0.00,0.07,1.00}{\quad\quad \leq I(X_1, X_2; Y| X_3) - I(X_1, X_2; Z) + I(X_1; Y| X_2, X_3) - I(X_1, X_3; Z) \leftarrow {\text {b3 + a4}}}\nonumber\\
& \textcolor[rgb]{1.00,0.00,0.00}{{\text {i10}}. \quad 2 R_1^{\text s} + R_2^{\text s} - 2 ( R_3^{\text o} + R_3^{\text g}) \leq I(X_1, X_2; Y| X_3) - I(X_1, X_3; Z) + I(X_1; Y| X_2, X_3) - I(X_1, X_2, X_3; Z),} \nonumber\\
& \textcolor[rgb]{0.00,0.07,1.00}{\quad\quad\quad [R_1^{\text s} - (R_3^{\text o} + R_3^{\text g})] + [(R_1^{\text s} + R_2^{\text s}) - (R_3^{\text o} + R_3^{\text g})] }\nonumber\\
& \textcolor[rgb]{0.00,0.07,1.00}{\quad\quad \leq I(X_1; Y| X_2, X_3) - I(X_1, X_3; Z) + I(X_1, X_2; Y| X_3) - I(X_1, X_2, X_3; Z) \leftarrow {\text {a4 + b5}}}\nonumber\\
& \textcolor[rgb]{1.00,0.00,0.00}{{\text {i11}}. \quad 2 R_1^{\text s} + R_2^{\text s} + R_3^{\text s} \leq I(X_1, X_2; Y| X_3) - I(X_1, X_3; Z) + I(X_1, X_3; Y| X_2) - I(X_1, X_2; Z),} \nonumber\\
& \textcolor[rgb]{0.00,0.07,1.00}{\quad\quad\quad (R_1^{\text s} + R_2^{\text s}) + ( R_1^{\text s} + R_3^{\text s})}\nonumber\\
& \textcolor[rgb]{0.00,0.07,1.00}{\quad\quad \leq I(X_1, X_2; Y| X_3) - I(X_1, X_2; Z) + I(X_1, X_3; Y| X_2) - I(X_1, X_3; Z) \leftarrow {\text {b3 + c4}}}\nonumber\\
& \textcolor[rgb]{1.00,0.00,0.00}{{\text {i12}}. \quad 2 R_1^{\text s} \!+\! R_2^{\text s} \!+\! R_3^{\text s} \!-\! (R_3^{\text o} \!+\! R_3^{\text g}) \!\leq\! I(X_1, X_2; Y| X_3) \!-\! I(X_1, X_3; Z) \!+\! I(X_1, X_3; Y| X_2) \!-\! I(X_1, X_2, X_3; Z),} \nonumber\\
& \textcolor[rgb]{0.00,0.07,1.00}{\quad\quad\quad ( R_1^{\text s} + R_2^{\text s} + R_3^{\text s}) + [R_1^{\text s} - (R_3^{\text o} + R_3^{\text g})] }\nonumber\\
& \textcolor[rgb]{0.00,0.07,1.00}{\quad\quad \leq I(X_1, X_2, X_3; Y) - I(X_1, X_2, X_3; Z) + I(X_1; Y| X_2, X_3) - I(X_1, X_3; Z) \leftarrow {\text {d5 + a4}}}\nonumber\\
& \textcolor[rgb]{0.00,0.07,1.00}{\quad\quad = I(X_2; Y) + I(X_1, X_3; Y| X_2) - I(X_1, X_2, X_3; Z) + I(X_1; Y| X_2, X_3) - I(X_1, X_3; Z) }\nonumber\\
& \textcolor[rgb]{0.00,0.07,1.00}{\quad\quad \leq I(X_1, X_2; Y| X_3) - I(X_1, X_3; Z) + I(X_1, X_3; Y| X_2) - I(X_1, X_2, X_3; Z)}\nonumber
\end{align}
\begin{align}
& {\text {j6}}. \quad R_1^{\text s} + R_1^{\text o} + R_2^{\text s} + R_2^{\text o} + R_3^{\text s} + R_3^{\text o} + R_3^{\text g} \leq I(X_1, X_2, X_3; Y), \nonumber\\
& {\text {j7}}. \quad R_1^{\text s} + R_1^{\text o} + R_2^{\text s} + R_3^{\text s} + R_3^{\text o} + R_3^{\text g} \leq I(X_1, X_2, X_3; Y) - I(X_2; Z), \nonumber\\
& \textcolor[rgb]{0.00,0.59,0.00}{{\text {j8}}. \quad R_1^{\text s} + R_1^{\text o} + R_2^{\text s} + R_3^{\text s} \leq I(X_1, X_2, X_3; Y) - I(X_2, X_3; Z),} \nonumber\\
& \textcolor[rgb]{1.00,0.00,0.00}{{\text {j9}}. \quad 2 R_1^{\text s} + R_1^{\text o} + R_2^{\text s} + R_3^{\text s} + R_3^{\text o} + R_3^{\text g} \leq I(X_1, X_2, X_3; Y) + I(X_1; Y| X_2, X_3) - I(X_1, X_2; Z),} \nonumber\\
& \textcolor[rgb]{0.00,0.07,1.00}{\quad\quad\quad (R_1^{\text s} + R_1^{\text o}) + (R_1^{\text s} + R_2^{\text s} + R_3^{\text s} + R_3^{\text o} + R_3^{\text g})}\nonumber\\
& \textcolor[rgb]{0.00,0.07,1.00}{\quad\quad \leq I(X_1; Y| X_2, X_3) + I(X_1, X_2, X_3; Y) - I(X_1, X_2; Z) \leftarrow {\text {a1 + d3}}}\nonumber\\
& \textcolor[rgb]{1.00,0.00,0.00}{{\text {j10}}. \quad 2 R_1^{\text s} + R_1^{\text o} + R_2^{\text s} + R_3^{\text s} \leq I(X_1, X_2, X_3; Y) + I(X_1; Y| X_2, X_3) - I(X_1, X_2, X_3; Z),} \nonumber\\
& \textcolor[rgb]{0.00,0.07,1.00}{\quad\quad\quad (R_1^{\text s} + R_1^{\text o}) + (R_1^{\text s} + R_2^{\text s} + R_3^{\text s})}\nonumber\\
& \textcolor[rgb]{0.00,0.07,1.00}{\quad\quad \leq I(X_1; Y| X_2, X_3) + I(X_1, X_2, X_3; Y) - I(X_1, X_2, X_3; Z) \leftarrow {\text {a1 + d5}} }\nonumber\\
& \textcolor[rgb]{1.00,0.00,0.00}{{\text {j11}}. \quad 2 R_1^{\text s} + R_1^{\text o} + R_2^{\text s} + 2 R_3^{\text s} + 2 R_3^{\text o} + 2 R_3^{\text g} \leq I(X_1, X_2, X_3; Y) + I(X_1, X_3; Y| X_2) - I(X_1, X_2; Z),} \nonumber\\
& \textcolor[rgb]{0.00,0.07,1.00}{\quad\quad\quad (R_1^{\text s} + R_2^{\text s} + R_3^{\text s} + R_3^{\text o} + R_3^{\text g}) + (R_1^{\text s} + R_1^{\text o} + R_3^{\text s} + R_3^{\text o} + R_3^{\text g})}\nonumber\\
& \textcolor[rgb]{0.00,0.07,1.00}{\quad\quad \leq I(X_1, X_2, X_3; Y) - I(X_1, X_2; Z) + I(X_1, X_3; Y| X_2) \leftarrow {\text {d3 + c1}} }\nonumber\\
& \textcolor[rgb]{1.00,0.00,0.00}{{\text {j12}}. \quad 2 R_1^{\text s} + R_1^{\text o} + R_2^{\text s} + 2 R_3^{\text s} + R_3^{\text o} + R_3^{\text g} \leq I(X_1, X_2, X_3; Y) + I(X_1, X_3; Y| X_2) - I(X_1, X_2, X_3; Z),} \nonumber\\
& \textcolor[rgb]{0.00,0.07,1.00}{\quad\quad\quad (R_1^{\text s} + R_2^{\text s} + R_3^{\text s}) + (R_1^{\text s} + R_1^{\text o} + R_3^{\text s} + R_3^{\text o} + R_3^{\text g})}\nonumber\\
& \textcolor[rgb]{0.00,0.07,1.00}{\quad\quad \leq I(X_1, X_2, X_3; Y) - I(X_1, X_2, X_3; Z) + I(X_1, X_3; Y| X_2) \leftarrow {\text {d5 + c1}} }\nonumber
\end{align}
\begin{align}
& {\text {k6}}. \quad R_1^{\text s} + R_2^{\text s} + R_2^{\text o} + R_3^{\text s} + R_3^{\text o} + R_3^{\text g} \leq I(X_1, X_2, X_3; Y) - I(X_1; Z), \nonumber\\
& \textcolor[rgb]{1.00,0.00,0.00}{{\text {k7}}. \quad R_1^{\text s} + R_2^{\text s} + R_3^{\text s} + R_3^{\text o} + R_3^{\text g} \leq I(X_1, X_2, X_3; Y) - I(X_1; Z) - I(X_2; Z),} \nonumber\\
& \textcolor[rgb]{0.00,0.07,1.00}{\quad\quad\quad R_1^{\text s} + R_2^{\text s} + R_3^{\text s} + R_3^{\text o} + R_3^{\text g} }\nonumber\\
& \textcolor[rgb]{0.00,0.07,1.00}{\quad\quad \leq I(X_1, X_2, X_3; Y) - I(X_1, X_2; Z) \leftarrow {\text {d3}}}\nonumber\\
& \textcolor[rgb]{0.00,0.07,1.00}{\quad\quad \leq I(X_1, X_2, X_3; Y) - I(X_1; Z) - I(X_2; Z)}\nonumber\\
& \textcolor[rgb]{1.00,0.00,0.00}{{\text {k8}}. \quad R_1^{\text s} + R_2^{\text s} + R_3^{\text s} \leq I(X_1, X_2, X_3; Y) - I(X_1; Z) - I(X_2, X_3; Z),} \nonumber\\
& \textcolor[rgb]{0.00,0.07,1.00}{\quad\quad\quad R_1^{\text s} + R_2^{\text s} + R_3^{\text s}}\nonumber\\
& \textcolor[rgb]{0.00,0.07,1.00}{\quad\quad \leq I(X_1, X_2, X_3; Y) - I(X_1, X_2, X_3; Z) \leftarrow {\text {d5}}}\nonumber\\
& \textcolor[rgb]{0.00,0.07,1.00}{\quad\quad \leq I(X_1, X_2, X_3; Y) - I(X_1; Z) - I(X_2, X_3; Z)}\nonumber\\
& \textcolor[rgb]{1.00,0.00,0.00}{{\text {k9}}. \quad 2 R_1^{\text s} + R_2^{\text s} + R_3^{\text s} + R_3^{\text o} + R_3^{\text g} \leq I(X_1, X_2, X_3; Y) - I(X_1; Z) + I(X_1; Y| X_2, X_3) - I(X_1, X_2; Z),} \nonumber\\
& \textcolor[rgb]{0.00,0.07,1.00}{\quad\quad\quad (R_1^{\text s}) + (R_1^{\text s} + R_2^{\text s} + R_3^{\text s} + R_3^{\text o} + R_3^{\text g})}\nonumber\\
& \textcolor[rgb]{0.00,0.07,1.00}{\quad\quad \leq I(X_1; Y| X_2, X_3) - I(X_1; Z) + I(X_1, X_2, X_3; Y) - I(X_1, X_2; Z) \leftarrow {\text {a2 + d3}}}\nonumber\\
& \textcolor[rgb]{1.00,0.00,0.00}{{\text {k10}}. \quad 2 R_1^{\text s} + R_2^{\text s} + R_3^{\text s} \leq I(X_1, X_2, X_3; Y) - I(X_1; Z) + I(X_1; Y| X_2, X_3) - I(X_1, X_2, X_3; Z),} \nonumber\\
& \textcolor[rgb]{0.00,0.07,1.00}{\quad\quad\quad (R_1^{\text s}) + (R_1^{\text s} + R_2^{\text s} + R_3^{\text s})}\nonumber\\
& \textcolor[rgb]{0.00,0.07,1.00}{\quad\quad \leq I(X_1; Y| X_2, X_3) - I(X_1; Z) + I(X_1, X_2, X_3; Y) - I(X_1, X_2, X_3; Z) \leftarrow {\text {a2 + d5}}}\nonumber\\
& \textcolor[rgb]{1.00,0.00,0.00}{{\text {k11}}. \quad 2 R_1^{\text s} + R_2^{\text s} + 2 R_3^{\text s} + 2 R_3^{\text o} + 2 R_3^{\text g} \leq I(X_1, X_2, X_3; Y) - I(X_1; Z) + I(X_1, X_3; Y| X_2) - I(X_1, X_2; Z),} \nonumber\\
& \textcolor[rgb]{0.00,0.07,1.00}{\quad\quad\quad (R_1^{\text s} + R_2^{\text s} + R_3^{\text s} + R_3^{\text o} + R_3^{\text g}) + (R_1^{\text s} + R_3^{\text s} + R_3^{\text o} + R_3^{\text g})}\nonumber\\
& \textcolor[rgb]{0.00,0.07,1.00}{\quad\quad \leq I(X_1, X_2, X_3; Y) - I(X_1, X_2; Z) + I(X_1, X_3; Y| X_2) - I(X_1; Z) \leftarrow {\text {d3 + c2}} }\nonumber\\
& \textcolor[rgb]{1.00,0.00,0.00}{{\text {k12}}. \quad 2 R_1^{\text s} + R_2^{\text s} + 2 R_3^{\text s} + R_3^{\text o} + R_3^{\text g} \leq I(X_1, X_2, X_3; Y) - I(X_1; Z) + I(X_1, X_3; Y| X_2) - I(X_1, X_2, X_3; Z),} \nonumber\\
& \textcolor[rgb]{0.00,0.07,1.00}{\quad\quad\quad (R_1^{\text s} + R_2^{\text s} + R_3^{\text s}) + (R_1^{\text s} + R_3^{\text s} + R_3^{\text o} + R_3^{\text g})}\nonumber\\
& \textcolor[rgb]{0.00,0.07,1.00}{\quad\quad \leq I(X_1, X_2, X_3; Y) - I(X_1, X_2, X_3; Z) + I(X_1, X_3; Y| X_2) - I(X_1; Z) \leftarrow {\text {d5 + c2}} }\nonumber
\end{align}
\begin{align}
& \textcolor[rgb]{0.00,0.59,0.00}{{\text {L6}}. \quad R_1^{\text s} + R_2^{\text s} + R_2^{\text o} + R_3^{\text s} \leq I(X_1, X_2, X_3; Y) - I(X_1, X_3; Z),} \nonumber\\
& \textcolor[rgb]{1.00,0.00,0.00}{{\text {L7}}. \quad R_1^{\text s} + R_2^{\text s} + R_3^{\text s} \leq I(X_1, X_2, X_3; Y) - I(X_1, X_3; Z) - I(X_2; Z),} \nonumber\\
& \textcolor[rgb]{0.00,0.07,1.00}{\quad\quad R_1^{\text s} + R_2^{\text s} + R_3^{\text s} \leq I(X_1, X_2, X_3; Y) - I(X_1, X_2, X_3; Z) \leftarrow {\text {d5}}}\nonumber\\
& \textcolor[rgb]{0.00,0.07,1.00}{\quad\quad\quad\quad\quad\quad\quad\quad\; \leq I(X_1, X_2, X_3; Y) - I(X_1, X_3; Z) - I(X_2; Z)}\nonumber\\
& \textcolor[rgb]{1.00,0.00,0.00}{{\text {L8}}. \quad R_1^{\text s} + R_2^{\text s} + R_3^{\text s} - (R_3^{\text o} + R_3^{\text g}) \leq I(X_1, X_2, X_3; Y) - I(X_1, X_3; Z) - I(X_2, X_3; Z),} \nonumber\\
& \textcolor[rgb]{0.00,0.07,1.00}{\quad\quad\quad (R_1^{\text s} + R_2^{\text s} + R_3^{\text s}) - (R_3^{\text o} + R_3^{\text g})}\nonumber\\
& \textcolor[rgb]{0.00,0.07,1.00}{\quad\quad \leq I(X_1, X_2, X_3; Y) - I(X_1, X_2, X_3; Z) - I(X_3; Z) \leftarrow {\text {d5 + (15)}}}\nonumber\\
& \textcolor[rgb]{0.00,0.07,1.00}{\quad\quad = I(X_1, X_2, X_3; Y) - I(X_1, X_3; Z) - I(X_2; Z| X_1, X_3) - I(X_3; Z) }\nonumber\\
& \textcolor[rgb]{0.00,0.07,1.00}{\quad\quad \leq I(X_1, X_2, X_3; Y) - I(X_1, X_3; Z) - I(X_2, X_3; Z)}\nonumber\\
& \textcolor[rgb]{1.00,0.00,0.00}{{\text {L9}}. \quad 2 R_1^{\text s} + R_2^{\text s} + R_3^{\text s} \leq I(X_1, X_2, X_3; Y) - I(X_1, X_3; Z) + I(X_1; Y| X_2, X_3) - I(X_1, X_2; Z),} \nonumber\\
& \textcolor[rgb]{0.00,0.07,1.00}{\quad\quad\quad (R_1^{\text s}) + (R_1^{\text s} + R_2^{\text s} + R_3^{\text s}) }\nonumber\\
& \textcolor[rgb]{0.00,0.07,1.00}{\quad\quad \leq I(X_1; Y| X_2, X_3) - I(X_1; Z) + I(X_1, X_2, X_3; Y) - I(X_1, X_2, X_3; Z) \leftarrow {\text {a2 + d5}}}\nonumber\\
& \textcolor[rgb]{0.00,0.07,1.00}{\quad\quad = I(X_1; Y| X_2, X_3) - I(X_1; Z) + I(X_1, X_2, X_3; Y) - I(X_1, X_3; Z) - I(X_2; Z| X_1, X_3) }\nonumber\\
& \textcolor[rgb]{0.00,0.07,1.00}{\quad\quad \leq I(X_1, X_2, X_3; Y) - I(X_1, X_3; Z) + I(X_1; Y| X_2, X_3) - I(X_1, X_2; Z)}\nonumber\\
& \textcolor[rgb]{1.00,0.00,0.00}{{\text {L10}}. \quad 2 R_1^{\text s} \!+\! R_2^{\text s} \!+\! R_3^{\text s} \!-\! (R_3^{\text o} \!+\! R_3^{\text g}) \!\leq\! I(X_1, X_2, X_3; Y) \!-\! I(X_1, X_3; Z) \!+\! I(X_1; Y| X_2, X_3) \!-\! I(X_1, X_2, X_3; Z),} \nonumber\\
& \textcolor[rgb]{0.00,0.07,1.00}{\quad\quad\quad (R_1^{\text s} + R_2^{\text s} + R_3^{\text s}) + [R_1^{\text s} - (R_3^{\text o} + R_3^{\text g})]}\nonumber\\
& \textcolor[rgb]{0.00,0.07,1.00}{\quad\quad \leq I(X_1, X_2, X_3; Y) - I(X_1, X_2, X_3; Z) + I(X_1; Y| X_2, X_3) - I(X_1, X_3; Z) \leftarrow {\text {d5 + a4}} }\nonumber\\
& \textcolor[rgb]{1.00,0.00,0.00}{{\text {L11}}. \quad 2 R_1^{\text s} + R_2^{\text s} + 2 R_3^{\text s} + R_3^{\text o} + R_3^{\text g} \leq I(X_1, X_2, X_3; Y) - I(X_1, X_3; Z) + I(X_1, X_3; Y| X_2) - I(X_1, X_2; Z),} \nonumber\\
& \textcolor[rgb]{0.00,0.07,1.00}{\quad\quad\quad (R_1^{\text s} + R_2^{\text s} + R_3^{\text s} + R_3^{\text o} + R_3^{\text g}) + (R_1^{\text s} + R_3^{\text s})}\nonumber\\
& \textcolor[rgb]{0.00,0.07,1.00}{\quad\quad \leq I(X_1, X_2, X_3; Y) - I(X_1, X_2; Z) + I(X_1, X_3; Y| X_2) - I(X_1, X_3; Z) \leftarrow {\text {d3 + c4}} }\nonumber\\
& \textcolor[rgb]{1.00,0.00,0.00}{{\text {L12}}. \quad 2 R_1^{\text s} + R_2^{\text s} + 2 R_3^{\text s} \leq I(X_1, X_2, X_3; Y) - I(X_1, X_3; Z) + I(X_1, X_3; Y| X_2) - I(X_1, X_2, X_3; Z),} \nonumber\\
& \textcolor[rgb]{0.00,0.07,1.00}{\quad\quad\quad (R_1^{\text s} + R_2^{\text s} + R_3^{\text s}) + (R_1^{\text s} + R_3^{\text s})}\nonumber\\
& \textcolor[rgb]{0.00,0.07,1.00}{\quad\quad \leq I(X_1, X_2, X_3; Y) - I(X_1, X_2, X_3; Z) + I(X_1, X_3; Y| X_2) - I(X_1, X_3; Z) \leftarrow {\text {d5 + c4}} }\nonumber
\end{align}

\subsection{Elimination of $R_3^{g}$}

From (\ref{region_before_elimi_1}) $\sim$ (\ref{region_before_elimi_17}) and the inequalities resulted from eliminating $R_1^{\text g}$ and $R_2^{\text g}$ in the previous two subsections, we get $9$ upper bounds on $R_3^{\text g}$
\begin{align}
(\ref{region_before_elimi_3}) \rightarrow & {\text m}. \quad R_3^{\text g} \leq I(X_3; Y| X_1, X_2) - (R_3^{\text s} + R_3^{\text o}), \nonumber\\
{\text {c1}} \rightarrow & {\text n}. \quad R_3^{\text g} \leq I(X_1, X_3; Y| X_2) - (R_1^{\text s} + R_1^{\text o} + R_3^{\text s} + R_3^{\text o}), \nonumber\\
{\text {c2}} \rightarrow & {\text o}. \quad R_3^{\text g} \leq I(X_1, X_3; Y| X_2) - I(X_1; Z) - (R_1^{\text s} + R_3^{\text s} + R_3^{\text o}), \nonumber\\
{\text {d3}} \rightarrow & {\text p}. \quad R_3^{\text g} \leq I(X_1, X_2, X_3; Y) - I(X_1, X_2; Z) - (R_1^{\text s} + R_2^{\text s} + R_3^{\text s} + R_3^{\text o}), \nonumber\\
{\text {f6}} \rightarrow & {\text q}. \quad R_3^{\text g} \leq I(X_2, X_3; Y| X_1) - (R_2^{\text s} + R_2^{\text o} + R_3^{\text s} + R_3^{\text o}), \nonumber\\
{\text {f7}} \rightarrow & {\text r}. \quad R_3^{\text g} \leq I(X_2, X_3; Y| X_1) - I(X_2; Z) - (R_2^{\text s} + R_3^{\text s} + R_3^{\text o}), \nonumber\\
{\text {j6}} \rightarrow & {\text s}. \quad R_3^{\text g} \leq I(X_1, X_2, X_3; Y) - (R_1^{\text s} + R_1^{\text o} + R_2^{\text s} + R_2^{\text o} + R_3^{\text s} + R_3^{\text o}), \nonumber\\
{\text {j7}} \rightarrow & {\text t}. \quad R_3^{\text g} \leq I(X_1, X_2, X_3; Y) - I(X_2; Z) - (R_1^{\text s} + R_1^{\text o} + R_2^{\text s} + R_3^{\text s} + R_3^{\text o}), \nonumber\\
{\text {k6}} \rightarrow & {\text u}. \quad R_3^{\text g} \leq I(X_1, X_2, X_3; Y) - I(X_1; Z) - (R_1^{\text s} + R_2^{\text s} + R_2^{\text o} + R_3^{\text s} + R_3^{\text o}), \nonumber
\end{align}
and $7$ lower bounds on $R_3^{\text g}$
\begin{align}
(\ref{region_before_elimi_10}) \rightarrow & {\text 13}. \quad R_3^{\text g} \geq 0, \nonumber\\
(\ref{region_before_elimi_13}) \rightarrow & {\text 14}. \quad R_3^{\text g} \geq I(X_3; Z) - R_3^{\text o}, \nonumber\\
{\text {a4}} \rightarrow & {\text 15}. \quad R_3^{\text g} \geq - I(X_1; Y| X_2, X_3) + I(X_1, X_3; Z) + R_1^{\text s} - R_3^{\text o}, \nonumber\\
{\text {b5}} \rightarrow & {\text 16}. \quad R_3^{\text g} \geq - I(X_1, X_2; Y| X_3) + I(X_1, X_2, X_3; Z) + R_1^{\text s} + R_2^{\text s} - R_3^{\text o}, \nonumber\\
{\text {e8}} \rightarrow & {\text 17}. \quad R_3^{\text g} \geq - I(X_2; Y| X_1, X_3) + I(X_2, X_3; Z) + R_2^{\text s} - R_3^{\text o}, \nonumber\\
{\text {g8}} \rightarrow & {\text 18}. \quad R_3^{\text g} \geq - I(X_1, X_2; Y| X_3) + I(X_2, X_3; Z) + R_1^{\text s} + R_1^{\text o} + R_2^{\text s} - R_3^{\text o}, \nonumber\\
{\text {i6}} \rightarrow & {\text 19}. \quad R_3^{\text g} \geq - I(X_1, X_2; Y| X_3) + I(X_1, X_3; Z) + R_1^{\text s} + R_2^{\text s} + R_2^{\text o} - R_3^{\text o}, \nonumber
\end{align}

Comparing the upper bounds ${\text m} \sim {\text u}$ and lower bounds $13 \sim 19$ on $R_3^{\text g}$ given above, we get
\begin{align}
& \textcolor[rgb]{0.00,0.59,0.00}{{\text {m13}}. \quad R_3^{\text s} + R_3^{\text o} \leq I(X_3; Y| X_1, X_2),} \nonumber\\
& \textcolor[rgb]{0.00,0.59,0.00}{{\text {m14}}. \quad R_3^{\text s} \leq I(X_3; Y| X_1, X_2) - I(X_3; Z),} \nonumber\\
& \textcolor[rgb]{1.00,0.00,0.00}{{\text {m15}}. \quad R_1^{\text s} + R_3^{\text s} \leq I(X_3; Y| X_1, X_2) + I(X_1; Y| X_2, X_3) - I(X_1, X_3; Z),} \nonumber\\
& \textcolor[rgb]{0.00,0.07,1.00}{\quad\quad\quad R_1^{\text s} + R_3^{\text s} \leq I(X_1, X_3; Y| X_2) - I(X_1, X_3; Z) \leftarrow {\text {c4}}}\nonumber\\
& \textcolor[rgb]{0.00,0.07,1.00}{\quad\quad\quad\quad\quad\quad\quad\!\! \leq I(X_3; Y| X_1, X_2) + I(X_1; Y| X_2, X_3) - I(X_1, X_3; Z)}\nonumber\\
& \textcolor[rgb]{1.00,0.00,0.00}{{\text {m16}}. \quad R_1^{\text s} + R_2^{\text s} + R_3^{\text s} \leq I(X_3; Y| X_1, X_2) + I(X_1, X_2; Y| X_3) - I(X_1, X_2, X_3; Z),} \nonumber\\
& \textcolor[rgb]{0.00,0.07,1.00}{\quad\quad\quad R_1^{\text s} + R_2^{\text s} + R_3^{\text s} \leq I(X_1, X_2, X_3; Y) - I(X_1, X_2, X_3; Z) \leftarrow {\text {d5}}}\nonumber\\
& \textcolor[rgb]{0.00,0.07,1.00}{\quad\quad\quad\quad\quad\quad\quad\quad\quad \leq I(X_3; Y| X_1, X_2) + I(X_1, X_2; Y| X_3) - I(X_1, X_2, X_3; Z)}\nonumber\\
& \textcolor[rgb]{1.00,0.00,0.00}{{\text {m17}}. \quad R_2^{\text s} + R_3^{\text s} \leq I(X_3; Y| X_1, X_2) + I(X_2; Y| X_1, X_3) - I(X_2, X_3; Z),} \nonumber\\
& \textcolor[rgb]{0.00,0.07,1.00}{\quad\quad\quad R_2^{\text s} + R_3^{\text s} \leq I(X_2, X_3; Y| X_1) - I(X_2, X_3; Z) \leftarrow {\text {f8}}}\nonumber\\
& \textcolor[rgb]{0.00,0.07,1.00}{\quad\quad\quad\quad\quad\quad\quad\!\! \leq I(X_3; Y| X_1, X_2) + I(X_2; Y| X_1, X_3) - I(X_2, X_3; Z)}\nonumber\\
& \textcolor[rgb]{1.00,0.00,0.00}{{\text {m18}}. \quad R_1^{\text s} + R_1^{\text o} + R_2^{\text s} + R_3^{\text s} \leq I(X_3; Y| X_1, X_2) + I(X_1, X_2; Y| X_3) - I(X_2, X_3; Z),} \nonumber\\
& \textcolor[rgb]{0.00,0.07,1.00}{\quad\quad\quad R_1^{\text s} + R_1^{\text o} + R_2^{\text s} + R_3^{\text s} \leq I(X_1, X_2, X_3; Y) - I(X_2, X_3; Z) \leftarrow {\text {j8}}}\nonumber\\
& \textcolor[rgb]{0.00,0.07,1.00}{\quad\quad\quad\quad\quad\quad\quad\quad\quad\quad\quad\quad\!\!\! \leq I(X_3; Y| X_1, X_2) + I(X_1, X_2; Y| X_3) - I(X_2, X_3; Z)}\nonumber\\
& \textcolor[rgb]{1.00,0.00,0.00}{{\text {m19}}. \quad R_1^{\text s} + R_2^{\text s} + R_2^{\text o} + R_3^{\text s} \leq I(X_3; Y| X_1, X_2) + I(X_1, X_2; Y| X_3) - I(X_1, X_3; Z),} \nonumber\\
& \textcolor[rgb]{0.00,0.07,1.00}{\quad\quad\quad R_1^{\text s} + R_2^{\text s} + R_2^{\text o} + R_3^{\text s} \leq I(X_1, X_2, X_3; Y) - I(X_1, X_3; Z) \leftarrow {\text {L6}}}\nonumber\\
& \textcolor[rgb]{0.00,0.07,1.00}{\quad\quad\quad\quad\quad\quad\quad\quad\quad\quad\quad\quad\!\!\! \leq I(X_3; Y| X_1, X_2) + I(X_1, X_2; Y| X_3) - I(X_1, X_3; Z)}\nonumber\\
& \textcolor[rgb]{0.00,0.07,1.00}{\quad\quad\quad {\text {or}}}\nonumber\\
& \textcolor[rgb]{0.00,0.07,1.00}{\quad\quad\quad (R_1^{\text s} + R_2^{\text s} + R_2^{\text o} ) + (R_3^{\text s}) }\nonumber\\
& \textcolor[rgb]{0.00,0.07,1.00}{\quad\quad \leq I(X_1, X_2; Y| X_3) - I(X_1; Z) + I(X_3; Y| X_1, X_2) - I(X_3; Z) \leftarrow {\text {h6 + m14}}}\nonumber\\
& \textcolor[rgb]{0.00,0.07,1.00}{\quad\quad \leq I(X_3; Y| X_1, X_2) + I(X_1, X_2; Y| X_3) - I(X_1, X_3; Z) }\nonumber
\end{align}
\begin{align}
& \textcolor[rgb]{0.00,0.59,0.00}{{\text {n13}}. \quad R_1^{\text s} + R_1^{\text o} + R_3^{\text s} + R_3^{\text o} \leq I(X_1, X_3; Y| X_2),} \nonumber\\
& \textcolor[rgb]{0.00,0.59,0.00}{{\text {n14}}. \quad R_1^{\text s} + R_1^{\text o} + R_3^{\text s} \leq I(X_1, X_3; Y| X_2) - I(X_3; Z),} \nonumber\\
& \textcolor[rgb]{1.00,0.00,0.00}{{\text {n15}}. \quad 2 R_1^{\text s} + R_1^{\text o} + R_3^{\text s} \leq I(X_1, X_3; Y| X_2) + I(X_1; Y| X_2, X_3) - I(X_1, X_3; Z),} \nonumber\\
& \textcolor[rgb]{0.00,0.07,1.00}{\quad\quad\quad (R_1^{\text s} + R_1^{\text o}) + (R_1^{\text s} + R_3^{\text s}) \leq I(X_1; Y| X_2, X_3) + I(X_1, X_3; Y| X_2) - I(X_1, X_3; Z) \leftarrow {\text {a1 + c4}}} \nonumber\\
& \textcolor[rgb]{0.00,0.07,1.00}{\quad\quad\quad {\text {or}} } \nonumber\\
& \textcolor[rgb]{0.00,0.07,1.00}{\quad\quad\quad (R_1^{\text s}) + (R_1^{\text s} + R_1^{\text o} + R_3^{\text s}) } \nonumber\\
& \textcolor[rgb]{0.00,0.07,1.00}{\quad\quad \leq I(X_1; Y| X_2, X_3) - I(X_1; Z) + I(X_1, X_3; Y| X_2) - I(X_3; Z) \leftarrow {\text {a2 + n14}} } \nonumber\\
& \textcolor[rgb]{0.00,0.07,1.00}{\quad\quad \leq I(X_1, X_3; Y| X_2) + I(X_1; Y| X_2, X_3) - I(X_1, X_3; Z) } \nonumber\\
& \textcolor[rgb]{1.00,0.00,0.00}{{\text {n16}}. \quad 2 R_1^{\text s} + R_1^{\text o} + R_2^{\text s} + R_3^{\text s} \leq I(X_1, X_3; Y| X_2) + I(X_1, X_2; Y| X_3) - I(X_1, X_2, X_3; Z),} \nonumber\\
& \textcolor[rgb]{0.00,0.07,1.00}{\quad\quad\quad (R_1^{\text s} + R_1^{\text o}) + (R_1^{\text s} + R_2^{\text s} + R_3^{\text s})}\nonumber\\
& \textcolor[rgb]{0.00,0.07,1.00}{\quad\quad \leq I(X_1; Y| X_2, X_3) + I(X_1, X_2, X_3; Y) - I(X_1, X_2, X_3; Z) \leftarrow {\text {a1 + d5}}}\nonumber\\
& \textcolor[rgb]{0.00,0.07,1.00}{\quad\quad = I(X_1; Y| X_2, X_3) + I(X_3; Y) + I(X_1, X_2; Y| X_3) - I(X_1, X_2, X_3; Z)}\nonumber\\
& \textcolor[rgb]{0.00,0.07,1.00}{\quad\quad \leq I(X_1, X_3; Y| X_2) + I(X_1, X_2; Y| X_3) - I(X_1, X_2, X_3; Z)}\nonumber\\
& \textcolor[rgb]{1.00,0.00,0.00}{{\text {n17}}. \quad R_1^{\text s} + R_1^{\text o} + R_2^{\text s} + R_3^{\text s} \leq I(X_1, X_3; Y| X_2) + I(X_2; Y| X_1, X_3) - I(X_2, X_3; Z),} \nonumber\\
& \textcolor[rgb]{0.00,0.07,1.00}{\quad\quad\quad R_1^{\text s} + R_1^{\text o} + R_2^{\text s} + R_3^{\text s} \leq I(X_1, X_2, X_3; Y) - I(X_2, X_3; Z) \leftarrow {\text {j8}}}\nonumber\\
& \textcolor[rgb]{0.00,0.07,1.00}{\quad\quad\quad\quad\quad\quad\quad\quad\quad\quad\quad\;\; \leq I(X_1, X_3; Y| X_2) + I(X_2; Y| X_1, X_3) - I(X_2, X_3; Z)}\nonumber\\
& \textcolor[rgb]{1.00,0.00,0.00}{{\text {n18}}. \quad 2 R_1^{\text s} + 2 R_1^{\text o} + R_2^{\text s} + R_3^{\text s} \leq I(X_1, X_3; Y| X_2) + I(X_1, X_2; Y| X_3) - I(X_2, X_3; Z),} \nonumber\\
& \textcolor[rgb]{0.00,0.07,1.00}{\quad\quad\quad (R_1^{\text s} + R_1^{\text o}) + (R_1^{\text s} + R_1^{\text o} + R_2^{\text s} + R_3^{\text s})}\nonumber\\
& \textcolor[rgb]{0.00,0.07,1.00}{\quad\quad \leq I(X_1; Y| X_2, X_3) + I(X_1, X_2, X_3; Y) - I(X_2, X_3; Z) \leftarrow {\text {a1 + j8}} }\nonumber\\
& \textcolor[rgb]{0.00,0.07,1.00}{\quad\quad = I(X_1; Y| X_2, X_3) + I(X_3; Y) + I(X_1, X_2; Y| X_3) - I(X_2, X_3; Z)}\nonumber\\
& \textcolor[rgb]{0.00,0.07,1.00}{\quad\quad \leq I(X_1, X_3; Y| X_2) + I(X_1, X_2; Y| X_3) - I(X_2, X_3; Z)}\nonumber\\
& \textcolor[rgb]{1.00,0.00,0.00}{{\text {n19}}. \quad 2 R_1^{\text s} + R_1^{\text o} + R_2^{\text s} + R_2^{\text o} + R_3^{\text s} \leq I(X_1, X_3; Y| X_2) + I(X_1, X_2; Y| X_3) - I(X_1, X_3; Z),} \nonumber\\
& \textcolor[rgb]{0.00,0.07,1.00}{\quad\quad\quad (R_1^{\text s} + R_1^{\text o}) + (R_1^{\text s} + R_2^{\text s} + R_2^{\text o} + R_3^{\text s}) }\nonumber\\
& \textcolor[rgb]{0.00,0.07,1.00}{\quad\quad \leq I(X_1; Y| X_2, X_3) + I(X_1, X_2, X_3; Y) - I(X_1, X_3; Z) \leftarrow {\text {a1 + L6}} }\nonumber\\
& \textcolor[rgb]{0.00,0.07,1.00}{\quad\quad = I(X_1; Y| X_2, X_3) + I(X_3; Y) + I(X_1, X_2; Y| X_3) - I(X_1, X_3; Z)}\nonumber\\
& \textcolor[rgb]{0.00,0.07,1.00}{\quad\quad \leq I(X_1, X_3; Y| X_2) + I(X_1, X_2; Y| X_3) - I(X_1, X_3; Z)}\nonumber
\end{align}
\begin{align}
& \textcolor[rgb]{0.00,0.59,0.00}{{\text {o13}}. \quad R_1^{\text s} + R_3^{\text s} + R_3^{\text o} \leq I(X_1, X_3; Y| X_2) - I(X_1; Z),} \nonumber\\
& \textcolor[rgb]{1.00,0.00,0.00}{{\text {o14}}. \quad R_1^{\text s} + R_3^{\text s} \leq I(X_1, X_3; Y| X_2) - I(X_1; Z) - I(X_3; Z),} \nonumber\\
& \textcolor[rgb]{0.00,0.07,1.00}{\quad\quad\quad R_1^{\text s} + R_3^{\text s} \leq I(X_1, X_3; Y| X_2) - I(X_1, X_3; Z) \leftarrow {\text { c4}}}\nonumber\\
& \textcolor[rgb]{0.00,0.07,1.00}{\quad\quad\quad\quad\quad\quad\quad\!\! \leq I(X_1, X_3; Y| X_2) - I(X_1; Z) - I(X_3; Z) }\nonumber\\
& \textcolor[rgb]{1.00,0.00,0.00}{{\text {o15}}. \quad 2 R_1^{\text s} + R_3^{\text s} \leq I(X_1, X_3; Y| X_2) - I(X_1; Z) + I(X_1; Y| X_2, X_3) - I(X_1, X_3; Z),} \nonumber\\
& \textcolor[rgb]{0.00,0.07,1.00}{\quad\quad\quad (R_1^{\text s}) + (R_1^{\text s} + R_3^{\text s})} \nonumber\\
& \textcolor[rgb]{0.00,0.07,1.00}{\quad\quad \leq I(X_1; Y| X_2, X_3) - I(X_1; Z) + I(X_1, X_3; Y| X_2) - I(X_1, X_3; Z) \leftarrow {\text {a2 + c4}}} \nonumber\\
& \textcolor[rgb]{1.00,0.00,0.00}{{\text {o16}}. \quad 2 R_1^{\text s} + R_2^{\text s} + R_3^{\text s} \leq I(X_1, X_3; Y| X_2) - I(X_1; Z) + I(X_1, X_2; Y| X_3) - I(X_1, X_2, X_3; Z),} \nonumber\\
& \textcolor[rgb]{0.00,0.07,1.00}{\quad\quad\quad (R_1^{\text s}) + (R_1^{\text s} + R_2^{\text s} + R_3^{\text s})}\nonumber\\
& \textcolor[rgb]{0.00,0.07,1.00}{\quad\quad \leq I(X_1; Y| X_2, X_3) - I(X_1; Z) + I(X_1, X_2, X_3; Y) - I(X_1, X_2, X_3; Z) \leftarrow {\text {a2 + d5}}}\nonumber\\
& \textcolor[rgb]{0.00,0.07,1.00}{\quad\quad = I(X_1; Y| X_2, X_3) - I(X_1; Z) + I(X_3; Y) + I(X_1, X_2; Y| X_3) - I(X_1, X_2, X_3; Z)}\nonumber\\
& \textcolor[rgb]{0.00,0.07,1.00}{\quad\quad \leq I(X_1, X_3; Y| X_2) - I(X_1; Z) + I(X_1, X_2; Y| X_3) - I(X_1, X_2, X_3; Z)}\nonumber\\
& \textcolor[rgb]{1.00,0.00,0.00}{{\text {o17}}. \quad R_1^{\text s} + R_2^{\text s} + R_3^{\text s} \leq I(X_1, X_3; Y| X_2) - I(X_1; Z) + I(X_2; Y| X_1, X_3) - I(X_2, X_3; Z),} \nonumber\\
& \textcolor[rgb]{0.00,0.07,1.00}{\quad\quad\quad R_1^{\text s} + R_2^{\text s} + R_3^{\text s} \leq I(X_1, X_2, X_3; Y) - I(X_1, X_2, X_3; Z) \leftarrow {\text {d5}}}\nonumber\\
& \textcolor[rgb]{0.00,0.07,1.00}{\quad\quad\quad\quad\quad\quad\quad\quad\quad \leq I(X_1, X_3; Y| X_2) - I(X_1; Z) + I(X_2; Y| X_1, X_3) - I(X_2, X_3; Z)} \nonumber\\
& \textcolor[rgb]{1.00,0.00,0.00}{{\text {o18}}. \quad 2 R_1^{\text s} + R_1^{\text o} + R_2^{\text s} + R_3^{\text s} \leq I(X_1, X_3; Y| X_2) - I(X_1; Z) + I(X_1, X_2; Y| X_3) - I(X_2, X_3; Z),} \nonumber\\
& \textcolor[rgb]{0.00,0.07,1.00}{\quad\quad\quad (R_1^{\text s} + R_1^{\text o}) + (R_1^{\text s} + R_2^{\text s} + R_3^{\text s})}\nonumber\\
& \textcolor[rgb]{0.00,0.07,1.00}{\quad\quad \leq I(X_1; Y| X_2, X_3) + I(X_1, X_2, X_3; Y) - I(X_1, X_2, X_3; Z) \leftarrow {\text {a1 + d5}} }\nonumber\\
& \textcolor[rgb]{0.00,0.07,1.00}{\quad\quad = I(X_1; Y| X_2, X_3) + I(X_3; Y) - I(X_1; Z) + I(X_1, X_2; Y| X_3) - I(X_2, X_3; Z| X_1)}\nonumber\\
& \textcolor[rgb]{0.00,0.07,1.00}{\quad\quad \leq I(X_1, X_3; Y| X_2) - I(X_1; Z) + I(X_1, X_2; Y| X_3) - I(X_2, X_3; Z)}\nonumber\\
& \textcolor[rgb]{1.00,0.00,0.00}{{\text {o19}}. \quad 2 R_1^{\text s} + R_2^{\text s} + R_2^{\text o} + R_3^{\text s} \leq I(X_1, X_3; Y| X_2) - I(X_1; Z) + I(X_1, X_2; Y| X_3) - I(X_1, X_3; Z),} \nonumber\\
& \textcolor[rgb]{0.00,0.07,1.00}{\quad\quad\quad (R_1^{\text s}) + (R_1^{\text s} + R_2^{\text s} + R_2^{\text o} + R_3^{\text s}) }\nonumber\\
& \textcolor[rgb]{0.00,0.07,1.00}{\quad\quad \leq I(X_1; Y| X_2, X_3) - I(X_1; Z) + I(X_1, X_2, X_3; Y) - I(X_1, X_3; Z) \leftarrow {\text {a2 + L6}} }\nonumber\\
& \textcolor[rgb]{0.00,0.07,1.00}{\quad\quad = I(X_1; Y| X_2, X_3) - I(X_1; Z) + I(X_3; Y) + I(X_1, X_2; Y| X_3) - I(X_1, X_3; Z)}\nonumber\\
& \textcolor[rgb]{0.00,0.07,1.00}{\quad\quad \leq I(X_1, X_3; Y| X_2) - I(X_1; Z) + I(X_1, X_2; Y| X_3) - I(X_1, X_3; Z)}\nonumber
\end{align}
\begin{align}
& \textcolor[rgb]{0.00,0.59,0.00}{{\text {p13}}. \quad R_1^{\text s} + R_2^{\text s} + R_3^{\text s} + R_3^{\text o} \leq I(X_1, X_2, X_3; Y) - I(X_1, X_2; Z),} \nonumber\\
& \textcolor[rgb]{1.00,0.00,0.00}{{\text {p14}}. \quad R_1^{\text s} + R_2^{\text s} + R_3^{\text s} \leq I(X_1, X_2, X_3; Y) - I(X_1, X_2; Z) - I(X_3; Z),} \nonumber\\
& \textcolor[rgb]{0.00,0.07,1.00}{\quad\quad\quad R_1^{\text s} + R_2^{\text s} + R_3^{\text s} \leq I(X_1, X_2, X_3; Y) - I(X_1, X_2, X_3; Z) \leftarrow {\text {d5}}}\nonumber\\
& \textcolor[rgb]{0.00,0.07,1.00}{\quad\quad\quad\quad\quad\quad\quad\quad\quad \leq I(X_1, X_2, X_3; Y) - I(X_1, X_2; Z) - I(X_3; Z)}\nonumber\\
& \textcolor[rgb]{1.00,0.00,0.00}{{\text {p15}}. \quad 2 R_1^{\text s} + R_2^{\text s} + R_3^{\text s} \leq I(X_1, X_2, X_3; Y) - I(X_1, X_2; Z) + I(X_1; Y| X_2, X_3) - I(X_1, X_3; Z),} \nonumber\\
& \textcolor[rgb]{0.00,0.07,1.00}{\quad\quad\quad (R_1^{\text s}) + (R_1^{\text s} + R_2^{\text s} + R_3^{\text s}) } \nonumber\\
& \textcolor[rgb]{0.00,0.07,1.00}{\quad\quad \leq I(X_1; Y| X_2, X_3) - I(X_1; Z) + I(X_1, X_2, X_3; Y) - I(X_1, X_2, X_3; Z) \leftarrow {\text {a2 + d5}}} \nonumber\\
& \textcolor[rgb]{0.00,0.07,1.00}{\quad\quad = I(X_1; Y| X_2, X_3) - I(X_1; Z) + I(X_1, X_2, X_3; Y) - I(X_2; Z| X_1, X_3) - I(X_1, X_3; Z)} \nonumber\\
& \textcolor[rgb]{0.00,0.07,1.00}{\quad\quad \leq I(X_1; Y| X_2, X_3) - I(X_1, X_2; Z) + I(X_1, X_2, X_3; Y) - I(X_1, X_3; Z)} \nonumber\\
& \textcolor[rgb]{1.00,0.00,0.00}{{\text {p16}}. \quad 2 R_1^{\text s} + 2 R_2^{\text s} + R_3^{\text s} \leq I(X_1, X_2, X_3; Y) - I(X_1, X_2; Z) + I(X_1, X_2; Y| X_3) - I(X_1, X_2, X_3; Z),} \nonumber\\
& \textcolor[rgb]{0.00,0.07,1.00}{\quad\quad\quad (R_1^{\text s} + R_2^{\text s}) + (R_1^{\text s} + R_2^{\text s} + R_3^{\text s})}\nonumber\\
& \textcolor[rgb]{0.00,0.07,1.00}{\quad\quad \leq I(X_1, X_2; Y| X_3) - I(X_1, X_2; Z) + I(X_1, X_2, X_3; Y) - I(X_1, X_2, X_3; Z) \leftarrow {\text {b3 + d5}}}\nonumber\\
& \textcolor[rgb]{1.00,0.00,0.00}{{\text {p17}}. \quad R_1^{\text s} + 2 R_2^{\text s} + R_3^{\text s} \leq I(X_1, X_2, X_3; Y) - I(X_1, X_2; Z) + I(X_2; Y| X_1, X_3) - I(X_2, X_3; Z),} \nonumber\\
& \textcolor[rgb]{0.00,0.07,1.00}{\quad\quad\quad (R_2^{\text s}) + (R_1^{\text s} + R_2^{\text s} + R_3^{\text s}) } \nonumber\\
& \textcolor[rgb]{0.00,0.07,1.00}{\quad\quad \leq I(X_2; Y| X_1, X_3) - I(X_2; Z) + I(X_1, X_2, X_3; Y) - I(X_1, X_2, X_3; Z) \leftarrow {\text {e7 + d5}}} \nonumber\\
& \textcolor[rgb]{0.00,0.07,1.00}{\quad\quad = I(X_2; Y| X_1, X_3) - I(X_2; Z) + I(X_1, X_2, X_3; Y) - I(X_1; Z| X_2, X_3) - I(X_2, X_3; Z)} \nonumber\\
& \textcolor[rgb]{0.00,0.07,1.00}{\quad\quad \leq I(X_2; Y| X_1, X_3) - I(X_1, X_2; Z) + I(X_1, X_2, X_3; Y) - I(X_2, X_3; Z)} \nonumber\\
& \textcolor[rgb]{1.00,0.00,0.00}{{\text {p18}}. \quad 2 R_1^{\text s} + R_1^{\text o} + 2 R_2^{\text s} + R_3^{\text s} \leq I(X_1, X_2, X_3; Y) - I(X_1, X_2; Z) + I(X_1, X_2; Y| X_3) - I(X_2, X_3; Z),} \nonumber\\
& \textcolor[rgb]{0.00,0.07,1.00}{\quad\quad\quad (R_1^{\text s} + R_1^{\text o} + R_2^{\text s}) + (R_1^{\text s} + R_2^{\text s} + R_3^{\text s})}\nonumber\\
& \textcolor[rgb]{0.00,0.07,1.00}{\quad\quad \leq I(X_1, X_2; Y| X_3) - I(X_2; Z) + I(X_1, X_2, X_3; Y) - I(X_1, X_2, X_3; Z)  \leftarrow {\text {g7 + d5}}}\nonumber\\
& \textcolor[rgb]{0.00,0.07,1.00}{\quad\quad = I(X_1, X_2, X_3; Y) - I(X_2; Z) - I(X_1; Z| X_2, X_3) + I(X_1, X_2; Y| X_3) - I(X_2, X_3; Z) }\nonumber\\
& \textcolor[rgb]{0.00,0.07,1.00}{\quad\quad \leq I(X_1, X_2, X_3; Y) - I(X_1, X_2; Z) + I(X_1, X_2; Y| X_3) - I(X_2, X_3; Z) }\nonumber\\
& \textcolor[rgb]{1.00,0.00,0.00}{{\text {p19}}. \quad 2 R_1^{\text s} + 2 R_2^{\text s} + R_2^{\text o} + R_3^{\text s} \leq I(X_1, X_2, X_3; Y) - I(X_1, X_2; Z) + I(X_1, X_2; Y| X_3) - I(X_1, X_3; Z),} \nonumber\\
& \textcolor[rgb]{0.00,0.07,1.00}{\quad\quad\quad (R_1^{\text s} + R_2^{\text s} + R_2^{\text o}) + (R_1^{\text s} + R_2^{\text s} + R_3^{\text s}) }\nonumber\\
& \textcolor[rgb]{0.00,0.07,1.00}{\quad\quad \leq I(X_1, X_2; Y| X_3) - I(X_1; Z) + I(X_1, X_2, X_3; Y) - I(X_1, X_2, X_3; Z)  \leftarrow {\text {h6 + d5}}}\nonumber\\
& \textcolor[rgb]{0.00,0.07,1.00}{\quad\quad \leq I(X_1, X_2, X_3; Y) - I(X_1, X_2; Z) + I(X_1, X_2; Y| X_3) - I(X_1, X_3; Z) }\nonumber
\end{align}
\begin{align}
& \textcolor[rgb]{0.00,0.59,0.00}{{\text {q13}}. \quad R_2^{\text s} + R_2^{\text o} + R_3^{\text s} + R_3^{\text o} \leq I(X_2, X_3; Y| X_1),} \nonumber\\
& \textcolor[rgb]{0.00,0.59,0.00}{{\text {q14}}. \quad R_2^{\text s} + R_2^{\text o} + R_3^{\text s} \leq I(X_2, X_3; Y| X_1) - I(X_3; Z),} \nonumber\\
& \textcolor[rgb]{1.00,0.00,0.00}{{\text {q15}}. \quad R_1^{\text s} + R_2^{\text s} + R_2^{\text o} + R_3^{\text s} \leq I(X_2, X_3; Y| X_1) + I(X_1; Y| X_2, X_3) - I(X_1, X_3; Z),} \nonumber\\
& \textcolor[rgb]{0.00,0.07,1.00}{\quad\quad\quad R_1^{\text s} + R_2^{\text s} + R_2^{\text o} + R_3^{\text s} \leq I(X_1, X_2, X_3; Y) - I(X_1, X_3; Z) \leftarrow {\text {L6}}}\nonumber\\
& \textcolor[rgb]{0.00,0.07,1.00}{\quad\quad\quad\quad\quad\quad\quad\quad\quad\quad\quad\;\; \leq I(X_2, X_3; Y| X_1) + I(X_1; Y| X_2, X_3) - I(X_1, X_3; Z)}\nonumber\\
& \textcolor[rgb]{1.00,0.00,0.00}{{\text {q16}}. \quad R_1^{\text s} + 2 R_2^{\text s} + R_2^{\text o} + R_3^{\text s} \leq I(X_2, X_3; Y| X_1) + I(X_1, X_2; Y| X_3) - I(X_1, X_2, X_3; Z),} \nonumber\\
& \textcolor[rgb]{0.00,0.07,1.00}{\quad\quad\quad (R_2^{\text s} + R_2^{\text o}) + (R_1^{\text s} + R_2^{\text s} + R_3^{\text s})}\nonumber\\
& \textcolor[rgb]{0.00,0.07,1.00}{\quad\quad \leq I(X_2; Y| X_1, X_3) + I(X_1, X_2, X_3; Y) - I(X_1, X_2, X_3; Z) \leftarrow {\text {e6 + d5}}}\nonumber\\
& \textcolor[rgb]{0.00,0.07,1.00}{\quad\quad \leq I(X_2, X_3; Y| X_1) + I(X_1, X_2; Y| X_3) - I(X_1, X_2, X_3; Z)}\nonumber\\
& \textcolor[rgb]{1.00,0.00,0.00}{{\text {q17}}. \quad 2 R_2^{\text s} + R_2^{\text o} + R_3^{\text s} \leq I(X_2, X_3; Y| X_1) + I(X_2; Y| X_1, X_3) - I(X_2, X_3; Z),} \nonumber\\
& \textcolor[rgb]{0.00,0.07,1.00}{\quad\quad\quad (R_2^{\text s} + R_2^{\text o}) + ( R_2^{\text s} + R_3^{\text s}) } \nonumber\\
& \textcolor[rgb]{0.00,0.07,1.00}{\quad\quad \leq I(X_2; Y| X_1, X_3) + I(X_2, X_3; Y| X_1) - I(X_2, X_3; Z) \leftarrow {\text {e6 + f8}}} \nonumber\\
& \textcolor[rgb]{1.00,0.00,0.00}{{\text {q18}}. \quad R_1^{\text s} + R_1^{\text o} + 2 R_2^{\text s} + R_2^{\text o} + R_3^{\text s} \leq I(X_2, X_3; Y| X_1) + I(X_1, X_2; Y| X_3) - I(X_2, X_3; Z),} \nonumber\\
& \textcolor[rgb]{0.00,0.07,1.00}{\quad\quad\quad (R_2^{\text s} + R_2^{\text o}) + (R_1^{\text s} + R_1^{\text o} + R_2^{\text s} + R_3^{\text s})}\nonumber\\
& \textcolor[rgb]{0.00,0.07,1.00}{\quad\quad \leq I(X_2; Y| X_1, X_3) + I(X_1, X_2, X_3; Y| X_1) - I(X_2, X_3; Z) \leftarrow {\text {e6 + j8}} }\nonumber\\
& \textcolor[rgb]{0.00,0.07,1.00}{\quad\quad \leq I(X_2, X_3; Y| X_1) + I(X_1, X_2; Y| X_3) - I(X_2, X_3; Z) }\nonumber\\
& \textcolor[rgb]{1.00,0.00,0.00}{{\text {q19}}. \quad R_1^{\text s} + 2 R_2^{\text s} + 2 R_2^{\text o} + R_3^{\text s} \leq I(X_2, X_3; Y| X_1) + I(X_1, X_2; Y| X_3) - I(X_1, X_3; Z),} \nonumber\\
& \textcolor[rgb]{0.00,0.07,1.00}{\quad\quad\quad (R_2^{\text s} + R_2^{\text o}) + (R_1^{\text s} + R_2^{\text s} + R_2^{\text o} + R_3^{\text s}) }\nonumber\\
& \textcolor[rgb]{0.00,0.07,1.00}{\quad\quad \leq I(X_2; Y| X_1, X_3) + I(X_1, X_2, X_3; Y) - I(X_1, X_3; Z) \leftarrow {\text {e6 + L6}} }\nonumber\\
& \textcolor[rgb]{0.00,0.07,1.00}{\quad\quad \leq I(X_2, X_3; Y| X_1) + I(X_1, X_2; Y| X_3) - I(X_1, X_3; Z) }\nonumber\\
& \textcolor[rgb]{0.00,0.07,1.00}{\quad\quad\quad {\text {or}}}\nonumber\\
& \textcolor[rgb]{0.00,0.07,1.00}{\quad\quad\quad (R_1^{\text s} + R_2^{\text s} + R_2^{\text o}) + (R_2^{\text s} + R_2^{\text o} + R_3^{\text s}) }\nonumber\\
& \textcolor[rgb]{0.00,0.07,1.00}{\quad\quad \leq I(X_1, X_2; Y| X_3) - I(X_1; Z) + I(X_2, X_3; Y| X_1) - I(X_3; Z) \leftarrow {\text {h6 + q14}} }\nonumber\\
& \textcolor[rgb]{0.00,0.07,1.00}{\quad\quad \leq I(X_2, X_3; Y| X_1) + I(X_1, X_2; Y| X_3) - I(X_1, X_3; Z) }\nonumber
\end{align}
\begin{align}
& \textcolor[rgb]{0.00,0.59,0.00}{{\text {r13}}. \quad R_2^{\text s} + R_3^{\text s} + R_3^{\text o} \leq I(X_2, X_3; Y| X_1) - I(X_2, Z),} \nonumber\\
& \textcolor[rgb]{1.00,0.00,0.00}{{\text {r14}}. \quad R_2^{\text s} + R_3^{\text s} \leq I(X_2, X_3; Y| X_1) - I(X_2, Z) - I(X_3; Z),} \nonumber\\
& \textcolor[rgb]{0.00,0.07,1.00}{\quad\quad\quad R_2^{\text s} + R_3^{\text s} \leq I(X_2, X_3; Y| X_1) - I(X_2, X_3; Z) \leftarrow {\text {f8}}}\nonumber\\
& \textcolor[rgb]{0.00,0.07,1.00}{\quad\quad\quad\quad\quad\quad\quad \leq I(X_2, X_3; Y| X_1) - I(X_2, Z) - I(X_3; Z)}\nonumber\\
& \textcolor[rgb]{1.00,0.00,0.00}{{\text {r15}}. \quad R_1^{\text s} + R_2^{\text s} + R_3^{\text s} \leq I(X_2, X_3; Y| X_1) - I(X_2, Z) + I(X_1; Y| X_2, X_3) - I(X_1, X_3; Z),} \nonumber\\
& \textcolor[rgb]{0.00,0.07,1.00}{\quad\quad\quad R_1^{\text s} + R_2^{\text s} + R_3^{\text s} \leq I(X_1, X_2, X_3; Y) - I(X_1, X_2, X_3; Z) \leftarrow {\text {d5}}}\nonumber\\
& \textcolor[rgb]{0.00,0.07,1.00}{\quad\quad\quad\quad\quad\quad\quad\quad\quad \leq I(X_2, X_3; Y| X_1) + I(X_1; Y| X_2, X_3) - I(X_2, Z) - I(X_1, X_3; Z)}\nonumber\\
& \textcolor[rgb]{1.00,0.00,0.00}{{\text {r16}}. \quad R_1^{\text s} + 2 R_2^{\text s} + R_3^{\text s} \leq I(X_2, X_3; Y| X_1) - I(X_2, Z) + I(X_1, X_2; Y| X_3) - I(X_1, X_2, X_3; Z),} \nonumber\\
& \textcolor[rgb]{0.00,0.07,1.00}{\quad\quad\quad (R_2^{\text s}) + (R_1^{\text s} + R_2^{\text s} + R_3^{\text s})}\nonumber\\
& \textcolor[rgb]{0.00,0.07,1.00}{\quad\quad \leq I(X_2; Y| X_1, X_3) - I(X_2, Z) + I(X_1, X_2, X_3; Y) - I(X_1, X_2, X_3; Z) \leftarrow {\text {e7 + d5}}}\nonumber\\
& \textcolor[rgb]{0.00,0.07,1.00}{\quad\quad \leq I(X_2, X_3; Y| X_1) - I(X_2, Z) + I(X_1, X_2; Y| X_3) - I(X_1, X_2, X_3; Z)}\nonumber\\
& \textcolor[rgb]{1.00,0.00,0.00}{{\text {r17}}. \quad 2 R_2^{\text s} + R_3^{\text s} \leq I(X_2, X_3; Y| X_1) - I(X_2, Z) + I(X_2; Y| X_1, X_3) - I(X_2, X_3; Z),} \nonumber\\
& \textcolor[rgb]{0.00,0.07,1.00}{\quad\quad\quad (R_2^{\text s}) + ( R_2^{\text s} + R_3^{\text s}) } \nonumber\\
& \textcolor[rgb]{0.00,0.07,1.00}{\quad\quad \leq I(X_2; Y| X_1, X_3) - I(X_2, Z) + I(X_2, X_3; Y| X_1) - I(X_2, X_3; Z) \leftarrow {\text {e7 + f8}}} \nonumber\\
& \textcolor[rgb]{1.00,0.00,0.00}{{\text {r18}}. \quad R_1^{\text s} + R_1^{\text o} + 2 R_2^{\text s} + R_3^{\text s} \leq I(X_2, X_3; Y| X_1) - I(X_2, Z) + I(X_1, X_2; Y| X_3) - I(X_2, X_3; Z),} \nonumber\\
& \textcolor[rgb]{0.00,0.07,1.00}{\quad\quad\quad (R_2^{\text s}) + (R_1^{\text s} + R_1^{\text o} + R_2^{\text s} + R_3^{\text s}) }\nonumber\\
& \textcolor[rgb]{0.00,0.07,1.00}{\quad\quad \leq I(X_2; Y| X_1, X_3) - I(X_2, Z) + I(X_1, X_2, X_3; Y) - I(X_2, X_3; Z) \leftarrow {\text {e7 + j8}} }\nonumber\\
& \textcolor[rgb]{0.00,0.07,1.00}{\quad\quad \leq I(X_2, X_3; Y| X_1) - I(X_2, Z) + I(X_1, X_2; Y| X_3) - I(X_2, X_3; Z) }\nonumber\\
& \textcolor[rgb]{1.00,0.00,0.00}{{\text {r19}}. \quad R_1^{\text s} + 2 R_2^{\text s} + R_2^{\text o} + R_3^{\text s} \leq I(X_2, X_3; Y| X_1) - I(X_2, Z) + I(X_1, X_2; Y| X_3) - I(X_1, X_3; Z),} \nonumber\\
& \textcolor[rgb]{0.00,0.07,1.00}{\quad\quad\quad (R_2^{\text s} + R_2^{\text o}) + (R_1^{\text s} + R_2^{\text s} + R_3^{\text s}) }\nonumber\\
& \textcolor[rgb]{0.00,0.07,1.00}{\quad\quad \leq I(X_2; Y| X_1, X_3) + I(X_1, X_2, X_3; Y) - I(X_1, X_2, X_3; Z) \leftarrow {\text {e6 + d5}} }\nonumber\\
& \textcolor[rgb]{0.00,0.07,1.00}{\quad\quad \leq I(X_2, X_3; Y| X_1) - I(X_2, Z) + I(X_1, X_2; Y| X_3) - I(X_1, X_3; Z) }\nonumber
\end{align}
\begin{align}
& \textcolor[rgb]{0.00,0.59,0.00}{{\text {s13}}. \quad R_1^{\text s} + R_1^{\text o} + R_2^{\text s} + R_2^{\text o} + R_3^{\text s} + R_3^{\text o} \leq I(X_1, X_2, X_3; Y),} \nonumber\\
& \textcolor[rgb]{0.00,0.59,0.00}{{\text {s14}}. \quad R_1^{\text s} + R_1^{\text o} + R_2^{\text s} + R_2^{\text o} + R_3^{\text s} \leq I(X_1, X_2, X_3; Y) - I(X_3; Z),} \nonumber\\
& \textcolor[rgb]{1.00,0.00,0.00}{{\text {s15}}. \quad 2 R_1^{\text s} + R_1^{\text o} + R_2^{\text s} + R_2^{\text o} + R_3^{\text s} \leq I(X_1, X_2, X_3; Y) + I(X_1; Y| X_2, X_3) - I(X_1, X_3; Z),} \nonumber\\
& \textcolor[rgb]{0.00,0.07,1.00}{\quad\quad\quad (R_1^{\text s} + R_1^{\text o}) + (R_1^{\text s} + R_2^{\text s} + R_2^{\text o} + R_3^{\text s})} \nonumber\\
& \textcolor[rgb]{0.00,0.07,1.00}{\quad\quad \leq I(X_1; Y| X_2, X_3) + I(X_1, X_2, X_3; Y) - I(X_1, X_3; Z) \leftarrow {\text {a1 + L6}} }\nonumber\\
& \textcolor[rgb]{1.00,0.00,0.00}{{\text {s16}}. \quad 2 R_1^{\text s} + R_1^{\text o} + 2 R_2^{\text s} + R_2^{\text o} + R_3^{\text s} \leq I(X_1, X_2, X_3; Y) + I(X_1, X_2; Y| X_3) - I(X_1, X_2, X_3; Z),} \nonumber\\
& \textcolor[rgb]{0.00,0.07,1.00}{\quad\quad\quad (R_1^{\text s} + R_1^{\text o} + R_2^{\text s} + R_2^{\text o}) + (R_1^{\text s} + R_2^{\text s} + R_3^{\text s})}\nonumber\\
& \textcolor[rgb]{0.00,0.07,1.00}{\quad\quad \leq I(X_1, X_2; Y| X_3) + I(X_1, X_2, X_3; Y) - I(X_1, X_2, X_3; Z) \leftarrow {\text {g6 + d5}}}\nonumber\\
& \textcolor[rgb]{1.00,0.00,0.00}{{\text {s17}}. \quad R_1^{\text s} + R_1^{\text o} + 2 R_2^{\text s} + R_2^{\text o} + R_3^{\text s} \leq I(X_1, X_2, X_3; Y) + I(X_2; Y| X_1, X_3) - I(X_2, X_3; Z),} \nonumber\\
& \textcolor[rgb]{0.00,0.07,1.00}{\quad\quad\quad (R_2^{\text s} + R_2^{\text o}) + (R_1^{\text s} + R_1^{\text o} + R_2^{\text s} + R_3^{\text s}) } \nonumber\\
& \textcolor[rgb]{0.00,0.07,1.00}{\quad\quad \leq I(X_2; Y| X_1, X_3) + I(X_1, X_2, X_3; Y) - I(X_2, X_3; Z) \leftarrow {\text {e6 + j8}} } \nonumber\\
& \textcolor[rgb]{1.00,0.00,0.00}{{\text {s18}}. \quad 2 R_1^{\text s} + 2 R_1^{\text o} + 2 R_2^{\text s} + R_2^{\text o} + R_3^{\text s} \leq I(X_1, X_2, X_3; Y) + I(X_1, X_2; Y| X_3) - I(X_2, X_3; Z),} \nonumber\\
& \textcolor[rgb]{0.00,0.07,1.00}{\quad\quad\quad (R_1^{\text s} + R_1^{\text o} + R_2^{\text s} + R_2^{\text o}) + (R_1^{\text s} + R_1^{\text o} + R_2^{\text s} + R_3^{\text s})}\nonumber\\
& \textcolor[rgb]{0.00,0.07,1.00}{\quad\quad \leq I(X_1, X_2; Y| X_3) + I(X_1, X_2, X_3; Y) - I(X_2, X_3; Z) \leftarrow {\text {g6 + j8}} }\nonumber\\
& \textcolor[rgb]{1.00,0.00,0.00}{{\text {s19}}. \quad 2 R_1^{\text s} + R_1^{\text o} + 2 R_2^{\text s} + 2 R_2^{\text o} + R_3^{\text s} \leq I(X_1, X_2, X_3; Y) + I(X_1, X_2; Y| X_3) - I(X_1, X_3; Z),} \nonumber\\
& \textcolor[rgb]{0.00,0.07,1.00}{\quad\quad\quad (R_1^{\text s} + R_1^{\text o} + R_2^{\text s} + R_2^{\text o}) + (R_1^{\text s} + R_2^{\text s} + R_2^{\text o} + R_3^{\text s}) }\nonumber\\
& \textcolor[rgb]{0.00,0.07,1.00}{\quad\quad \leq I(X_1, X_2; Y| X_3) + I(X_1, X_2, X_3; Y) - I(X_1, X_3; Z) \leftarrow {\text {g6 + L6}} }\nonumber\\
& \textcolor[rgb]{0.00,0.07,1.00}{\quad\quad\quad {\text{or}} }\nonumber\\
& \textcolor[rgb]{0.00,0.07,1.00}{\quad\quad\quad (R_1^{\text s} + R_2^{\text s} + R_2^{\text o}) + (R_1^{\text s} + R_1^{\text o} + R_2^{\text s} + R_2^{\text o} + R_3^{\text s}) }\nonumber\\
& \textcolor[rgb]{0.00,0.07,1.00}{\quad\quad \leq I(X_1, X_2; Y| X_3) - I(X_1; Z) + I(X_1, X_2, X_3; Y) - I(X_3; Z) \leftarrow {\text {h6 + s14}} }\nonumber\\
& \textcolor[rgb]{0.00,0.07,1.00}{\quad\quad \leq I(X_1, X_2, X_3; Y) + I(X_1, X_2; Y| X_3) - I(X_1, X_3; Z) }\nonumber
\end{align}
\begin{align}
& \textcolor[rgb]{0.00,0.59,0.00}{{\text {t13}}. \quad R_1^{\text s} + R_1^{\text o} + R_2^{\text s} + R_3^{\text s} + R_3^{\text o} \leq I(X_1, X_2, X_3; Y) - I(X_2; Z),} \nonumber\\
& \textcolor[rgb]{1.00,0.00,0.00}{{\text {t14}}. \quad R_1^{\text s} + R_1^{\text o} + R_2^{\text s} + R_3^{\text s} \leq I(X_1, X_2, X_3; Y) - I(X_2; Z) - I(X_3; Z),} \nonumber\\
& \textcolor[rgb]{0.00,0.07,1.00}{\quad\quad\quad R_1^{\text s} + R_1^{\text o} + R_2^{\text s} + R_3^{\text s} \leq I(X_1, X_2, X_3; Y) - I(X_2, X_3; Z) \leftarrow {\text {j8}}}\nonumber\\
& \textcolor[rgb]{0.00,0.07,1.00}{\quad\quad\quad\quad\quad\quad\quad\quad\quad\quad\quad\quad\!\!\! \leq I(X_1, X_2, X_3; Y) - I(X_2; Z) - I(X_3; Z)}\nonumber\\
& \textcolor[rgb]{1.00,0.00,0.00}{{\text {t15}}. \quad 2 R_1^{\text s} + R_1^{\text o} + R_2^{\text s} + R_3^{\text s} \leq I(X_1, X_2, X_3; Y) - I(X_2; Z) + I(X_1; Y| X_2, X_3) - I(X_1, X_3; Z),} \nonumber\\
& \textcolor[rgb]{0.00,0.07,1.00}{\quad\quad\quad (R_1^{\text s} + R_1^{\text o}) + (R_1^{\text s} + R_2^{\text s} + R_3^{\text s})} \nonumber\\
& \textcolor[rgb]{0.00,0.07,1.00}{\quad\quad \leq I(X_1; Y| X_2, X_3) + I(X_1, X_2, X_3; Y) - I(X_1, X_2, X_3; Z) \leftarrow {\text {a1 + d5}} }\nonumber\\
& \textcolor[rgb]{0.00,0.07,1.00}{\quad\quad \leq I(X_1; Y| X_2, X_3) + I(X_1, X_2, X_3; Y) - I(X_2; Z) - I(X_1, X_3; Z) }\nonumber\\
& \textcolor[rgb]{1.00,0.00,0.00}{{\text {t16}}. \quad 2 R_1^{\text s} + R_1^{\text o} + 2 R_2^{\text s} + R_3^{\text s} \leq I(X_1, X_2, X_3; Y) - I(X_2; Z) + I(X_1, X_2; Y| X_3) - I(X_1, X_2, X_3; Z),} \nonumber\\
& \textcolor[rgb]{0.00,0.07,1.00}{\quad\quad\quad (R_1^{\text s} + R_1^{\text o} + R_2^{\text s}) + (R_1^{\text s} + R_2^{\text s} + R_3^{\text s})}\nonumber\\
& \textcolor[rgb]{0.00,0.07,1.00}{\quad\quad \leq I(X_1, X_2; Y| X_3) - I(X_2; Z) + I(X_1, X_2, X_3; Y) - I(X_1, X_2, X_3; Z) \leftarrow {\text {g7 + d5}} }\nonumber\\
& \textcolor[rgb]{1.00,0.00,0.00}{{\text {t17}}. \quad R_1^{\text s} + R_1^{\text o} + 2 R_2^{\text s} + R_3^{\text s} \leq I(X_1, X_2, X_3; Y) - I(X_2; Z) + I(X_2; Y| X_1, X_3) - I(X_2, X_3; Z),} \nonumber\\
& \textcolor[rgb]{0.00,0.07,1.00}{\quad\quad\quad (R_2^{\text s}) + (R_1^{\text s} + R_1^{\text o} + R_2^{\text s} + R_3^{\text s}) } \nonumber\\
& \textcolor[rgb]{0.00,0.07,1.00}{\quad\quad \leq I(X_2; Y| X_1, X_3) - I(X_2; Z) + I(X_1, X_2, X_3; Y) - I(X_2, X_3; Z) \leftarrow {\text {e7 + j8}}} \nonumber\\
& \textcolor[rgb]{1.00,0.00,0.00}{{\text {t18}}. \quad 2 R_1^{\text s} + 2 R_1^{\text o} + 2 R_2^{\text s} + R_3^{\text s} \leq I(X_1, X_2, X_3; Y) - I(X_2; Z) + I(X_1, X_2; Y| X_3) - I(X_2, X_3; Z),} \nonumber\\
& \textcolor[rgb]{0.00,0.07,1.00}{\quad\quad\quad (R_1^{\text s} + R_1^{\text o} + R_2^{\text s}) + (R_1^{\text s} + R_1^{\text o} + R_2^{\text s} + R_3^{\text s})}\nonumber\\
& \textcolor[rgb]{0.00,0.07,1.00}{\quad\quad \leq I(X_1, X_2; Y| X_3) - I(X_2; Z) + I(X_1, X_2, X_3; Y) - I(X_2, X_3; Z) \leftarrow {\text {g7 + j8}} }\nonumber\\
& \textcolor[rgb]{1.00,0.00,0.00}{{\text {t19}}. \quad 2 R_1^{\text s} + R_1^{\text o} + 2 R_2^{\text s} + R_2^{\text o} + R_3^{\text s} \leq I(X_1, X_2, X_3; Y) - I(X_2; Z) + I(X_1, X_2; Y| X_3) - I(X_1, X_3; Z),} \nonumber\\
& \textcolor[rgb]{0.00,0.07,1.00}{\quad\quad\quad (R_1^{\text s} + R_1^{\text o} + R_2^{\text s} + R_2^{\text o}) + (R_1^{\text s} + R_2^{\text s} + R_3^{\text s}) }\nonumber\\
& \textcolor[rgb]{0.00,0.07,1.00}{\quad\quad \leq I(X_1, X_2; Y| X_3) + I(X_1, X_2, X_3; Y) - I(X_1, X_2, X_3; Z) \leftarrow {\text {g6 + d5}}}\nonumber\\
& \textcolor[rgb]{0.00,0.07,1.00}{\quad\quad \leq I(X_1, X_2, X_3; Y) - I(X_2; Z) + I(X_1, X_2; Y| X_3) - I(X_1, X_3; Z)}\nonumber\\
& \textcolor[rgb]{0.00,0.07,1.00}{\quad\quad\quad {\text{or}} }\nonumber\\
& \textcolor[rgb]{0.00,0.07,1.00}{\quad\quad\quad (R_1^{\text s} + R_1^{\text o} + R_2^{\text s}) + (R_1^{\text s} + R_2^{\text s} + R_2^{\text o} + R_3^{\text s}) }\nonumber\\
& \textcolor[rgb]{0.00,0.07,1.00}{\quad\quad \leq I(X_1, X_2; Y| X_3) - I(X_2; Z) + I(X_1, X_2, X_3; Y) - I(X_1, X_3; Z) \leftarrow {\text {g7 + L6}}}\nonumber
\end{align}
\begin{align}
& \textcolor[rgb]{0.00,0.59,0.00}{{\text {u13}}. \quad R_1^{\text s} + R_2^{\text s} + R_2^{\text o} + R_3^{\text s} + R_3^{\text o} \leq I(X_1, X_2, X_3; Y) - I(X_1; Z),} \nonumber\\
& \textcolor[rgb]{1.00,0.00,0.00}{{\text {u14}}. \quad R_1^{\text s} + R_2^{\text s} + R_2^{\text o} + R_3^{\text s} \leq I(X_1, X_2, X_3; Y) - I(X_1; Z) - I(X_3; Z),} \nonumber\\
& \textcolor[rgb]{0.00,0.07,1.00}{\quad\quad\quad R_1^{\text s} + R_2^{\text s} + R_2^{\text o} + R_3^{\text s} \leq I(X_1, X_2, X_3; Y) - I(X_1, X_3; Z) \leftarrow {\text {L6}}}\nonumber\\
& \textcolor[rgb]{0.00,0.07,1.00}{\quad\quad\quad\quad\quad\quad\quad\quad\quad\quad\quad\;\; \leq I(X_1, X_2, X_3; Y) - I(X_1; Z) - I(X_3; Z)}\nonumber\\
& \textcolor[rgb]{1.00,0.00,0.00}{{\text {u15}}. \quad 2 R_1^{\text s} + R_2^{\text s} + R_2^{\text o} + R_3^{\text s} \leq I(X_1, X_2, X_3; Y) - I(X_1; Z) + I(X_1; Y| X_2, X_3) - I(X_1, X_3; Z),} \nonumber\\
& \textcolor[rgb]{0.00,0.07,1.00}{\quad\quad\quad (R_1^{\text s}) + (R_1^{\text s} + R_2^{\text s} + R_2^{\text o} + R_3^{\text s})} \nonumber\\
& \textcolor[rgb]{0.00,0.07,1.00}{\quad\quad \leq I(X_1; Y| X_2, X_3) - I(X_1; Z) + I(X_1, X_2, X_3; Y) - I(X_1, X_3; Z) \leftarrow {\text {a2 + L6}} }\nonumber\\
& \textcolor[rgb]{1.00,0.00,0.00}{{\text {u16}}. \quad 2 R_1^{\text s} + 2 R_2^{\text s} + R_2^{\text o} + R_3^{\text s} \leq I(X_1, X_2, X_3; Y) - I(X_1; Z) + I(X_1, X_2; Y| X_3) - I(X_1, X_2, X_3; Z),} \nonumber\\
& \textcolor[rgb]{0.00,0.07,1.00}{\quad\quad\quad (R_1^{\text s} + R_2^{\text s} + R_2^{\text o}) + (R_1^{\text s} + R_2^{\text s} + R_3^{\text s})}\nonumber\\
& \textcolor[rgb]{0.00,0.07,1.00}{\quad\quad \leq I(X_1, X_2; Y| X_3) - I(X_1; Z) + I(X_1, X_2, X_3; Y) - I(X_1, X_2, X_3; Z) \leftarrow {\text {h6 + d5}} }\nonumber\\
& \textcolor[rgb]{1.00,0.00,0.00}{{\text {u17}}. \quad R_1^{\text s} + 2 R_2^{\text s} + R_2^{\text o} + R_3^{\text s} \leq I(X_1, X_2, X_3; Y) - I(X_1; Z) + I(X_2; Y| X_1, X_3) - I(X_2, X_3; Z),} \nonumber\\
& \textcolor[rgb]{0.00,0.07,1.00}{\quad\quad\quad (R_2^{\text s} + R_2^{\text o}) + (R_1^{\text s} + R_2^{\text s} + R_3^{\text s}) } \nonumber\\
& \textcolor[rgb]{0.00,0.07,1.00}{\quad\quad \leq I(X_2; Y| X_1, X_3) + I(X_1, X_2, X_3; Y) - I(X_1, X_2, X_3; Z) \leftarrow {\text {e6 + d5}} } \nonumber\\
& \textcolor[rgb]{0.00,0.07,1.00}{\quad\quad \leq I(X_1, X_2, X_3; Y) - I(X_1; Z) + I(X_2; Y| X_1, X_3) - I(X_2, X_3; Z) } \nonumber\\
& \textcolor[rgb]{1.00,0.00,0.00}{{\text {u18}}. \quad 2 R_1^{\text s} + R_1^{\text o} + 2 R_2^{\text s} + R_2^{\text o} + R_3^{\text s} \leq I(X_1, X_2, X_3; Y) - I(X_1; Z) + I(X_1, X_2; Y| X_3) - I(X_2, X_3; Z),} \nonumber\\
& \textcolor[rgb]{0.00,0.07,1.00}{\quad\quad\quad (R_1^{\text s} + R_1^{\text o} + R_2^{\text s} + R_2^{\text o}) + (R_1^{\text s} + R_2^{\text s} + R_3^{\text s})}\nonumber\\
& \textcolor[rgb]{0.00,0.07,1.00}{\quad\quad \leq I(X_1, X_2; Y| X_3) + I(X_1, X_2, X_3; Y) - I(X_1, X_2, X_3; Z) \leftarrow {\text {g6 + d5}} }\nonumber\\
& \textcolor[rgb]{0.00,0.07,1.00}{\quad\quad \leq I(X_1, X_2; Y| X_3) + I(X_1, X_2, X_3; Y) - I(X_1; Z) - I(X_2, X_3; Z) }\nonumber\\
& \textcolor[rgb]{1.00,0.00,0.00}{{\text {u19}}. \quad 2 R_1^{\text s} + 2 R_2^{\text s} + 2 R_2^{\text o} + R_3^{\text s} \leq I(X_1, X_2, X_3; Y) - I(X_1; Z) + I(X_1, X_2; Y| X_3) - I(X_1, X_3; Z),} \nonumber\\
& \textcolor[rgb]{0.00,0.07,1.00}{\quad\quad\quad (R_1^{\text s} + R_2^{\text s} + R_2^{\text o}) + (R_1^{\text s} + R_2^{\text s} + R_2^{\text o} + R_3^{\text s}) }\nonumber\\
& \textcolor[rgb]{0.00,0.07,1.00}{\quad\quad \leq I(X_1, X_2; Y| X_3) - I(X_1; Z) + I(X_1, X_2, X_3; Y) - I(X_1, X_3; Z) \leftarrow {\text {h6 + L6}} }\nonumber
\end{align}

\newpage
\subsection{Elimination Results}

By collecting all the efficient projected inequalities, which are in green color, we get the projection of (\ref{region_before_elimi_1}) $\sim$ (\ref{region_before_elimi_17}) onto the hyperplane $\{ R_1^{\text g} = 0, R_2^{\text g} = 0, R_3^{\text g} = 0\}$ as follows

\begin{equation}\label{projected_system}
\left\{\!\!\!
\begin{array}{ll}
\textcolor[rgb]{0.00,0.59,0.00}{{\text {a1}}. \quad R_1^{\text s} + R_1^{\text o} \leq I(X_1; Y| X_2, X_3)} \\
\textcolor[rgb]{0.00,0.59,0.00}{{\text {e6}}. \quad R_2^{\text s} + R_2^{\text o} \leq I(X_2; Y| X_1, X_3)}\\
\textcolor[rgb]{0.00,0.59,0.00}{{\text {m13}}. \quad R_3^{\text s} + R_3^{\text o} \leq I(X_3; Y| X_1, X_2)} \\
\textcolor[rgb]{0.00,0.59,0.00}{{\text {g6}}. \quad R_1^{\text s} + R_1^{\text o} + R_2^{\text s} + R_2^{\text o} \leq I(X_1, X_2; Y| X_3)}\\
\textcolor[rgb]{0.00,0.59,0.00}{{\text {n13}}. \quad R_1^{\text s} + R_1^{\text o} + R_3^{\text s} + R_3^{\text o} \leq I(X_1, X_3; Y| X_2)} \\
\textcolor[rgb]{0.00,0.59,0.00}{{\text {q13}}. \quad R_2^{\text s} + R_2^{\text o} + R_3^{\text s} + R_3^{\text o} \leq I(X_2, X_3; Y| X_1)} \\
\textcolor[rgb]{0.00,0.59,0.00}{{\text {s13}}. \quad R_1^{\text s} + R_1^{\text o} + R_2^{\text s} + R_2^{\text o} + R_3^{\text s} + R_3^{\text o} \leq I(X_1, X_2, X_3; Y)} \\
\textcolor[rgb]{0.00,0.59,0.00}{{\text {a2}}. \quad R_1^{\text s} \leq I(X_1; Y| X_2, X_3) - I(X_1; Z)} \\
\textcolor[rgb]{0.00,0.59,0.00}{{\text {e7}}. \quad R_2^{\text s} \leq I(X_2; Y| X_1, X_3) - I(X_2; Z)}\\
\textcolor[rgb]{0.00,0.59,0.00}{{\text {m14}}. \quad R_3^{\text s} \leq I(X_3; Y| X_1, X_2) - I(X_3; Z)} \\
\textcolor[rgb]{0.00,0.59,0.00}{{\text {b3}}. \quad R_1^{\text s} + R_2^{\text s} \leq I(X_1, X_2; Y| X_3) - I(X_1, X_2; Z)} \\
\textcolor[rgb]{0.00,0.59,0.00}{{\text {c4}}. \quad R_1^{\text s} + R_3^{\text s} \leq I(X_1, X_3; Y| X_2) - I(X_1, X_3; Z)}\\
\textcolor[rgb]{0.00,0.59,0.00}{{\text {f8}}. \quad R_2^{\text s} + R_3^{\text s} \leq I(X_2, X_3; Y| X_1) - I(X_2, X_3; Z)}\\
\textcolor[rgb]{0.00,0.59,0.00}{{\text {d5}}. \quad R_1^{\text s} + R_2^{\text s} + R_3^{\text s} \leq I(X_1, X_2, X_3; Y) - I(X_1, X_2, X_3; Z)}\\
\textcolor[rgb]{0.00,0.59,0.00}{{\text {g7}}. \quad R_1^{\text s} + R_1^{\text o} + R_2^{\text s} \leq I(X_1, X_2; Y| X_3) - I(X_2; Z)}\\
\textcolor[rgb]{0.00,0.59,0.00}{{\text {n14}}. \quad R_1^{\text s} + R_1^{\text o} + R_3^{\text s} \leq I(X_1, X_3; Y| X_2) - I(X_3; Z)} \\
\textcolor[rgb]{0.00,0.59,0.00}{{\text {h6}}. \quad R_1^{\text s} + R_2^{\text s} + R_2^{\text o} \leq I(X_1, X_2; Y| X_3) - I(X_1; Z)}\\
\textcolor[rgb]{0.00,0.59,0.00}{{\text {q14}}. \quad R_2^{\text s} + R_2^{\text o} + R_3^{\text s} \leq I(X_2, X_3; Y| X_1) - I(X_3; Z)} \\
\textcolor[rgb]{0.00,0.59,0.00}{{\text {o13}}. \quad R_1^{\text s} + R_3^{\text s} + R_3^{\text o} \leq I(X_1, X_3; Y| X_2) - I(X_1; Z)}\\
\textcolor[rgb]{0.00,0.59,0.00}{{\text {r13}}. \quad R_2^{\text s} + R_3^{\text s} + R_3^{\text o} \leq I(X_2, X_3; Y| X_1) - I(X_2, Z)}\\
\textcolor[rgb]{0.00,0.59,0.00}{{\text {j8}}. \quad R_1^{\text s} + R_1^{\text o} + R_2^{\text s} + R_3^{\text s} \leq I(X_1, X_2, X_3; Y) - I(X_2, X_3; Z)}\\
\textcolor[rgb]{0.00,0.59,0.00}{{\text {L6}}. \quad R_1^{\text s} + R_2^{\text s} + R_2^{\text o} + R_3^{\text s} \leq I(X_1, X_2, X_3; Y) - I(X_1, X_3; Z)}\\
\textcolor[rgb]{0.00,0.59,0.00}{{\text {p13}}. \quad R_1^{\text s} + R_2^{\text s} + R_3^{\text s} + R_3^{\text o} \leq I(X_1, X_2, X_3; Y) - I(X_1, X_2; Z)}\\
\textcolor[rgb]{0.00,0.59,0.00}{{\text {s14}}. \quad R_1^{\text s} + R_1^{\text o} + R_2^{\text s} + R_2^{\text o} + R_3^{\text s} \leq I(X_1, X_2, X_3; Y) - I(X_3; Z)}\\
\textcolor[rgb]{0.00,0.59,0.00}{{\text {t13}}. \quad R_1^{\text s} + R_1^{\text o} + R_2^{\text s} + R_3^{\text s} + R_3^{\text o} \leq I(X_1, X_2, X_3; Y) - I(X_2; Z)}\\
\textcolor[rgb]{0.00,0.59,0.00}{{\text {u13}}. \quad R_1^{\text s} + R_2^{\text s} + R_2^{\text o} + R_3^{\text s} + R_3^{\text o} \leq I(X_1, X_2, X_3; Y) - I(X_1; Z)}
\end{array} \right.
\end{equation}
which can be written as (\ref{rate_region0}) in short.
Theorem~\ref{theorem_FM} for the $K=3$ case is thus proven.

\section{Conclusions}
\label{conclusion}

In this note, we have proven Theorem~\ref{theorem_FM} for the $K=3$ case by directly applying Fourier-Motzkin elimination. 
Specifically, we eliminated $R_1^{\text g}$, $R_2^{\text g}$, and $R_3^{\text g}$ in (\ref{region_FM1}) one by one and showed that (\ref{rate_region0}) is a projection of (\ref{region_FM1}) onto hyperplane $\{ R_1^{\text g} = 0, R_2^{\text g} = 0, R_3^{\text g} = 0\}$.
The elimination procedure showed that so many inequalities are generated in the process and most them are redundant, indicating that it will become impractical to prove Theorem~\ref{theorem_FM} in this brute-force way if $K$ is large.
Therefore, we hope to strictly prove Theorem~\ref{theorem_FM} for the general case since it plays quite an important role for the capacity analysis of MAC-WT channels.
In our work \cite{xu2022achievable2}, we have provided the general proof in Appendix A.

\bibliographystyle{IEEEtran}
\bibliography{IEEEabrv,Ref}

\end{document}